\documentclass[12pt]{article}

\usepackage{epsfig}
\usepackage{amsmath}
\usepackage{amsfonts} 
\usepackage{graphicx}
\usepackage[german, english]{babel}
\usepackage{a4wide}
\usepackage{amsmath}
\usepackage{amssymb}
\usepackage{ifthen}
\usepackage{epsfig}

\date{\today}

\newcommand{\la}{\lambda}

\newcommand{\al}{\alpha}

\newcommand{\si}{\sigma}
\newcommand{\ta}{\theta}
\newcommand{\f}{\phi}
\newcommand{\vf}{\varphi}

\newcommand{\ee}{\end{equation}}
\newcommand{\eea}{\end{eqnarray}}
\newcommand{\be}{\begin{equation}}
\newcommand{\bea}{\begin{eqnarray}}

\newcommand{\pa}{\partial}

\newcommand{\vep}{\varepsilon}

\newcommand{\re}[1]{(\ref{#1})}
\newcommand{\R}{{\rm I \hspace{-0.52ex} R}}

\newcommand{\insertplot}[5]{\begin{figure}
 \hfill\hbox to 0.05in{\vbox to #5in{\vfill
 \inputplot{#1}{#4}{#5}}\hfill}
 \hfill\vspace{-.1in}
 \caption{#2}\label{#3}
 \end{figure}}
 \newcommand{\inputplot}[3]{
 \special{ps: plotfile #1}
}
\newcounter{fig}   

\begin{document}
\begin{center}

{\Large Goldstone models in $D+1$ dimensions, $D=3,4,5$, supporting 
\\
stable
and zero topological charge solutions}
\vspace{0.6cm}
\\
Eugen Radu$^{\dagger}$
and  D. H. Tchrakian$^{\dagger \star}$
\\
$^{\dagger  }${\small School of Theoretical Physics -- DIAS, 10 
Burlington
Road, Dublin 4, Ireland }
\\
$^{\star}${\small Department of
Mathematical Physics, National University of Ireland Maynooth,}
\\
\end{center}
\begin{abstract}
%%%%%%%%%%%%%%%%%%%%%
We study finite energy static solutions to a global symmetry breaking Goldstone
model described by an isovector scalar field in $D+1$ spacetime dimensions.
 Both topologically stable
multisolitons with arbitrary winding numbers, and zero topological 
charge soliton--antisoliton solutions
are constructed numerically in $D=3,4,5$. We have explored the types 
of symmetries the systems should
be subjected to, for there to exist multisoliton and 
soliton--antisoliton pairs in $D=3,4,5,6$. These
findings are underpinned by constructing numerical 
solutions in the $D\le 5$ examples. Subject to axial symmetry,
only multisolitons of all topological charges exist in {\it even} $D$, 
and in {\it odd} $D$, only zero and unit
topological charge solutions exist. Subjecting the system to weaker 
than axial symmetries, results
in the existence of all the possibilities in all dimensions. 
Our findings apply also to finite 'energy' solutions
to Yang--Mills and Yang-Mills--Higgs systems, and in principle also sigma models.
\end{abstract}

%%%%%%%%%%%%%%%%%%%%%%%%%%%%%%%%%%%%%%%%%%%%%%%%%%%%%%%%%%%%%%%%%%
\section{Introduction}
%%%%%%%%%%%%%%%%%%%%%%%%%%%%%%%%%%%%%%%%%%%%%%%%%%%%%%%%%%%%%%%%%%

Very early in the history of field theory solitons, interest in the existence of
zero topological charge solutions arose. In the case of the Yang--Mills (YM)
instantons~\cite{Belavin:1975fg} in $D=4$ Euclidean space, 
which are selfdual, this raised the question of
existence of non-selfdual~\cite{SS1,SS2,Bor:1992ai} solutions, while even earlier this
question was investigated~\cite{Taubes:1982ie} in the case of magnetic monopoles of the
YM-Higgs (YMH) model in $D=3$ \cite{mono}. More recently, concrete numerical constructions of
monopole--antimonopole solutions~\cite{R,KK,KKS} to the YMH model in $D=3$,
instanton--antiinstanton solutions~\cite{Radu:2006gg} to the YM model in $D=4$, as well as
soliton--antisoliton solutions~\cite{Paturyan:2005ik} to a Goldstone model in $D=3$, were given.

The potential relevance of field theory soliton--antisolitons in higher dimensions
rests in the fact that they describe non BPS field configurations that may be
useful in the description of brane--antibrane configurations. Non BPS configurations are
relevant for example in the context of string junctions in $N = 4$
super-Yang-Mills~\cite{Ioannidou:1999ca}.
Such solutions can be the zero topological charge counterparts
of higher dimensional instantons~\cite{Tchinst} and of monopoles~\cite{Tchmon}, or
of the solitons of the symmetry breaking Goldstone type models~\cite{gold-d} arising as
the gauge decoupling limits of higher dimensional monopole models~\cite{Tchmon}. These
Goldstone models have not found any physical applications to date, but as prototype
systems modeling higher dimensional monopoles without the burden of gauge degrees of
freedom, they can be useful for example in providing backgrounds on which Dirac
equations~\cite{dirac} in all dimensions can be solved, or possibly also for gravitating
monopoles. Here, they will prove very useful in studying zero topological charge
solutions in higher dimensions.

Zero topological charge solutions to such a Goldstone model in $D=3$ were recently
given in \cite{Paturyan:2005ik}. The model in \cite{Paturyan:2005ik}
is the gauge decoupling limit of the YMH model descending from the $p=2$ member of the YM
hierarchy introduced in \cite{Tchinst} and is the simplest example. In the present paper we will
extend this study to the Goldstone model descending from the $p=3$ member of the YM
hierarchy. In contrast to the $p=2$ Goldstone model which supports finite energy
solitons only in $D=3$,
the $p=3$ Goldstone model enables us to study solutions in dimensions $D=3,4,5$, allowed
by the Derrick scaling rule. This is very important for our purposes here as will be
explained below. As such, the $p=3$ Goldstone model will serve as a vehicle for us to
investigate zero topological charge solutions in the simplest possible technical setting.

The main objective of this work is to find out: subject to what symmetries and for
what boundary conditions do such solutions exist? We have presented several numerically
constructed solutions in dimensions $D\le 5$, by way of underpinning our findings.
While our symmetry considerations cover the dimensions $3\le D\le 6$, the concrete
numerical constructions are limited to $D=3,4,5$ in the $p=3$ Goldstone model,
covering both even and odd  dimensions, allowing us to make a classification of the said conditions. 
Our study addresses the question as to what are the requisite ingredients in the construction 
of zero topological charge solitons in higher dimensions, highlighting the distinction between 
even and odd dimensions in this respect. We find that such solutions can 
be accommodated by imposing the requisite boundary conditions for systems subject to the
appropriate symmetry, in all dimensions. Stated most succinctly, subject to axial symmetry only
multisolitons of arbitrary charges exist in even $D$, while in odd $D$ zero and unit topological
charge solutions can exist. By imposing less stringent symmetries than axial, all possible types of
solutions can be constructed in any dimension.

The symmetries considered are at one extreme {\it axial}, 
namely spherical symmetry in a $\R^{D-1}$
subspace of $\R^{D}$, and at the other {\it azimuthal}, 
namely rotational symmetry in a $\R^{2}$
subspace of $\R^{D}$. In between, we have explored the 
imposition of all {\it intermediate} cases,
namely imposition of rotational symmetry in all the other 
subspaces $\R^n$ of $\R^{D}$. In addition,
we have considered imposition of multi-azimuthal symmetries 
on all $\R^{2}$ subspaces of $\R^{D}$.

Concerning the numerical constructions, our reason for limiting 
to the $p=3$ Goldstone model, and to
$D\le 5$, is that otherwise it would be necessary to carry out 
numerical integrations in more than $2$
dimensions, which is beyond the scope of this work.

In Section  {\bf 2} we have introduced the models to be 
employed, along with the topological charge
densities providing the lower bounds on the energies. 
Section {\bf 3} is concerned with the imposition
of symmetries, {\it i.e.} stating the axial, azimuthal, 
intermediate and multi-azimuthal Ans\"atze. 
In particular, the energy density functionals of the model, 
for the dimensions in which numerical solutions
will be constructed, are subjected to the spherical, axial and bi-azimuthal symmetries.
Subjecting the corresponding topological charge densities to symmetries is carried out 
in Section {\bf 4}.
Section {\bf 5} contains all the numerical results, which
verify the assertions presented in the previous Section, {\bf 4}, 
concerning the symmetry properties
that zero topological charge solutions must have. 
In Section {\bf 6} we summarise our results and
extend the discussion of symmetries to beyond the particular simple models employed here.

%%%%%%%%%%%%%%%%%%%%%%%%%%%%%%%%%%%%%%%%%%%%%%%%%%%%%%%%%%%%%%%%%%%
\section{The model and the topological charge}
%%%%%%%%%%%%%%%%%%%%%%%%%%%%%%%%%%%%%%%%%%%%%%%%%%%%%%%%%%%%%%%%%%%
The symmetry breaking models in $D=3,~4$ and $5$ spatial dimensions, 
to which we
refer as Goldstone models, are described by a scalar isovector field 
$\f^a$,
$a=1,2,3$, $a=1,2,3,4$ and $a=1,2,3,4,5$ in each dimension, 
respectively.

There is such a hierarchy of models \cite{gold-d} that arise from 
the
gauge decoupled limit of the $D-$dimensional $SO(D)$ gauged Higgs 
(YMH) model
descended from the $p$-th member of the Yang-Mills (YM) hierarchy on
$\R_D\times S^{4p-D}$. Here we have chosen the simplest of these that 
can
accommodate $D=3,4,5$, while satisfying the Derrick scaling 
requirement for the
existence of finite energy solutions. In the present case this is the 
YMH model
that descends from the $p=3$-rd member of the YM hierarchy. Our 
Goldstone model
here is the gauge decoupled limit of this YMH system. Using the 
notation
\[
\f_{\mu}^a=\pa_\f{\mu}^a\quad,\quad\f_{\mu\nu}^{ab}=\pa_{[\mu}\f^a\pa_{\nu]}\f^b
\quad,\quad\f_{\mu\nu\rho}^{abc}=\pa_{[\mu}\f^a\pa_{\nu}\f^b\pa_{\rho]}\f^c\,,
\]
with the brackets $[\mu\nu...]$ implying total antisymmetrisation, 
the static
energy density is
\be
\label{en2}
{\cal E}_{(p=3)}=\la_0\,\left(\eta^2-|\f^a|^2\right)^6+\la_1\,
\left(\eta^2-|\f^b|^2\right)^4\,\left|\f_{\mu}^a\right|^2+
\la_2\left(\eta^2-|\f^b|^2\right)^2\left|\f_{\mu\nu}^{ab}\right|^2+
\la_3\left|\f_{\mu\nu\rho}^{abc}\right|^2\,.
\ee
All the dimensionless constants $\la_0$, $\la_1$, $\la_2$ and $\la_3$ 
must be
positive if the relevant topological lower bounds in each dimension 
are to be
valid.
%\footnote{By using a suitable rescaling, one can always set 
%$\lambda=1$,
%without any loss of generality.}. 
The model \re{en2} is {\it ad hoc} rather than a dimensionally 
descended,
only inasfar as the numerical values of these dimensionless coupling 
constants,
which are otherwise fixed by the descent mechanism, are constrained 
only to be
positive. Of course, any of these constants can also vanish, provided 
that the
absence of the corresponding term in \re{en2} does not violate the 
Derrick scaling
requirement.

The most important feature of models such as \re{en2} is that the 
order parameter
field $\f^a$ is a relic of a Higgs field and has the same dimensions 
($L^{-1}$) as
a connection, and the finite energy conditions allow the symmetry 
breaking
boundary conditions
\be
\label{bc}
\lim_{R\to 0}|\f^a|=0\quad,\quad\lim_{R\to\infty}|\f^a|=\eta\,,
\ee
$R$ being the radial coordinate in $\R^D$, this resulting in {\it monopole}-like
asymptotics for our solitons.

The presence of the symmetry breaking potential in \re{en2}, multiplying $\la_0$,
has no quantitative effect on the solutions, so it will be ignored henceforth.

In the next Section, where symmetries will be imposed, we will concentrate only on the terms
\be
\label{blocks}
\left|\f_{\mu}^{a}\right|^2\quad,\quad\left|\f_{\mu\nu}^{ab}\right|^2\quad,\quad
\left|\f_{\mu\nu\rho}^{abc}\right|^2
\ee
and will delay the incorporation of the factors $\left(\eta^2-|\f^b|^2\right)^2$ and
$\left(\eta^2-|\f^b|^2\right)^4$ till the Section on numerics, since imposition of
symmetries on these last terms is achieved rather trivially.

The topological charge density bounding the energy density functional 
from below can be stated simply in terms of Bogomol'nyi inequalities, separately for 
each dimension $D=3,4$ and $5$.

In $D=3$ the inequality
\be
\label{bog3}
\left(\eta^2-|\f|^2\right)^2\left|\left(\eta^2-|\f|^2\right)\f_{\rho}^c-
\frac{1}{2!^2}\vep_{\mu\nu\rho}\vep^{abc}\f_{\mu\nu}^{ab}\right|^2\ge 
0
\ee
leads to the lower bound~\footnote{Note that the $D=3$ model employed 
here is slightly
different from that in \cite{Paturyan:2005ik}, the latter being the 
gauge decoupled version of
the $p=2$ YMH model, in contrast to the gauge decoupled version of
the $p=3$ YMH model here.}
\be
\label{bogg3}
\left(\eta^2-|\f|^2\right)^4\left|\f_{\mu}^a\right|^2+\frac14
\left(\eta^2-|\f|^2\right)^2
\left|\f_{\mu\nu}^{ab}\right|^2\ge\vep_{\mu\nu\rho}\vep^{abc}
\left(\eta^2-|\f|^2\right)^3\f_{\mu}^a\f_{\nu}^b\f_{\rho}^c\equiv\varrho_3\,.
\ee
In \re{bog3}-\re{bogg3}, we have denoted $|\f^a|^2$ by $|\f|^2$.

In $D=4$ the inequalities
\bea
\nonumber
\left(\eta^2-|\f|^2\right)^2\left|\f_{\mu\nu}^{ab}-
\frac{1}{2!^2}\vep_{\mu\nu\rho\si}\vep^{abcd}\f_{\rho\si}^{cd}\right|^2&\ge&0
\label{bog4a}
\\
\left|\left(\eta^2-|\f|^2\right)^2\f_{\rho}^c-
\frac{1}{3!}\vep_{\mu\nu\rho\si}\vep^{abcd}
\f_{\mu\nu\rho}^{abc}\right|^2&\ge&0
\label{bog4b}
\eea
lead to the lower bound
\be
\label{bogg4}
\frac14\left(\eta^2-|\f|^2\right)^4\left|\f_{\mu}^a\right|^2+\frac12
\left(\eta^2-|\f|^2\right)^2\left|\f_{\mu\nu}^{ab}\right|^2+\frac14
\left|\f_{\mu\nu\rho}^{abc}\right|^2\ge\vep_{\mu\nu\rho\si}\vep^{abcd}
\left(\eta^2-|\f|^2\right)^2\f_{\mu}^a\f_{\nu}^b\f_{\rho}^c\f_{\si}^d\equiv\varrho_4\,.
\ee
\\
In $D=5$ the inequality
\bea
\nonumber
\label{bog5}
\left|\left(\eta^2-|\f|^2\right)\f_{\mu\nu}^{ab}-
\frac{1}{2^23!}\vep_{\mu\nu\rho\si\tau}\vep^{abcde}
\f_{\rho\si\tau}^{cde}\right|^2&\ge&0\label{bog5b}
\eea
leads to the lower bound
\be
\label{bogg5}
\left(\eta^2-|\f|^2\right)^2\left|\f_{\mu\nu}^{ab}\right|^2+\frac14
\left|\f_{\mu\nu\rho}^{abc}\right|^2\ge\vep_{\mu\nu\rho\si\tau}\vep^{abcde}
\left(\eta^2-|\f|^2\right)\f_{\mu}^a\f_{\nu}^b\f_{\rho}^c\f_{\si}^d\f_{\tau}^e\equiv
\varrho_5\,.
\ee

Each of the three topological charge densities $\varrho_3$, 
$\varrho_4$ and
$\varrho_5$ is a total divergence, which we denote as
$\varrho_3=\pa_{\mu}\Omega^{(3)}_{\mu}$, 
$\varrho_4=\pa_{\mu}\Omega^{(4)}_{\mu}$ and
$\varrho_5=\pa_{\mu}\Omega^{(5)}_{\mu}$, respectively, the surface 
integrals of
$\Omega^{(D)}_{\mu}$ yielding the topological charge in each 
dimension $D=3,4,5$. In
this paper we will refer to the densities $\Omega^{(D)}_{\mu}$ as 
{\it topological
currents}. Now these topological charges are simply numerical 
multiples of the
respective winding number densities
\be
\label{wnd}
\varrho_D^{(0)}=\vep_{\mu_1\mu_1...\mu_D}\,\vep^{a_1a_2...a_D}\,\f_{\mu_1}^{a_1}
\f_{\mu_2}^{a_2}...\f_{\mu_D}^{a_D}\equiv\pa_{\mu_1}\omega^{(D)}_{\mu_1}\,
\ee
which are the surface integrals of the winding number {\it currents}
\be
\label{wnf}
\omega^{(D)}_{\mu_1}=\vep_{\mu_1\mu_1...\mu_D}\,\vep^{a_1a_2...a_D}\,\f^{a_1}
\f_{\mu_2}^{a_2}...\f_{\mu_D}^{a_D}\,.
\ee

The topological charges $q_3,q_4$ and $q_5$, which are the volume 
integrals of the
densities $\varrho_3$, $\varrho_4$ and $\varrho_5$ defined in 
\re{bogg3}, \re{bogg4}
and \re{bogg5} respectively, are in turn equal to the surface 
integrals of the
topological currents
\bea
\nonumber
\Omega^{(3)}_{\mu}&=&\vep_{\mu\nu\rho}\vep^{abc}\left[\eta^6-\frac95\eta^4\left|
\f\right|^2+\frac97\eta^2\left(\left|\f\right|^2\right)^2-\frac13
\left(\left|\f\right|^2\right)^3\right]\,\f^a\f_{\nu}^b\f_{\rho}^c
\label{O3}
\\
\Omega^{(4)}_{\mu}&=&\vep_{\mu\nu\rho\si}\vep^{abcd}\left[\eta^4-\frac43\eta^2\left|
\f\right|^2+\frac12\eta^2\left(\left|\f\right|^2\right)^2\right]\,
\f^a\f_{\nu}^b\f_{\rho}^c\f_{\si}^d
\label{O4}
\\
\nonumber
\Omega^{(5)}_{\mu}&=&\vep_{\mu\nu\rho\si\tau}\vep^{abcde}
\left[\eta^2-\frac57\eta^2\left|\f\right|^2\right]\,
\f^a\f_{\nu}^b\f_{\rho}^c\f_{\si}^d\f_{\tau}^e\,.
\label{O5}
\eea
It is now obvious, in light of the asymptotic boundary value in 
\re{bc}, that
$q_3,q_4$ and $q_5$ are multiples of the winding numbers, namely the 
surface integrals
of the currents \re{wnf}, with the numbers
$\frac{16}{105}$, $\frac16$ and $\frac17$, respectively.

%%%%%%%%%%%%%%%%%%%%%%%%%%%%%%%%%%%%%%%%%%%%%%%%%%%%%%%%%%%%%%%%%%
\section{Imposition of symmetries}
%%%%%%%%%%%%%%%%%%%%%%%%%%%%%%%%%%%%%%%%%%%%%%%%%%%%%%%%%%%%%%%%%%

This section is divided into four subsections, in each of which the three
building blocks in \re{blocks} will be subjected to spherical, axial, azimuthal, and in $4$ dimensions
only, the bi-azimuthal symmetries respectively. We shall also state the tri-azimuthal Ansatz, only in
$6$ dimensions, but will not display the building blocks \re{blocks} subject to it because in
$6$ dimensions the Derrick scaling requires an {\it octic} term beyond these, whose numerical
pursuit is beyond the scope of this work. The
symmetry breaking selfinteraction potential $(\eta^2-|\f^a|^2)^6$ will be
ignored, instead the boundary condition \re{bc} it would enforce will be imposed directly.

Imposition of symmetry is the first step in the construction of zero topological charge solutions,
leading to the second step of selecting the requisite boundary conditions to achieve this aim. In this
section, we impose the symmetries on the energy density functional \re{en2} whose second order
equations will be integrated numerically in Section {\bf 5}, 
deferring the task of imposing symmetries
on the topological charge densities \re{wnd} and their currents \re{wnf} to the next section, {\bf 4}.
There, the most important task of selecting the requisite boundary conditions will be made.

Before stating the Ans\"atze pertaining to the various symmetries 
to be imposed on the scalar field $\f^a$ describing the model \re{en2}, we introduce the
coordinates to be employed in each case. Next to spherical symmetry, the strongest symmetry
that we will impose is the {\it axial} symmetry, sometimes described 
also as {\it cylindric} symmetry. This involves imposition of rotational symmetry in a $\R^{D-1}$
subspace of the full space $\R^{D}$. The weakest symmetry is the {\it azimuthal} one, which involves
imposition of rotational symmetry in a $\R^{2}$ subspace of $\R^{D}$. Then there are all the
{\it intermediate} symmetries involving imposition of rotational symmetry in a $\R^{n}$
subspace of $\R^{D}$, with $D-2\ge n\ge 3$. As we restrict to $D=5$, the only relevant
values of $n$ are $n=3$ and $4$. In addition, we will employ {\it multi-azimuthal}
symmetries, each one of its constituent azimuthal symmetries being imposed on distinct planes in $\R^D$.
Since we will restrict to $D=6$, our attention will be restricted to the {\it bi-azimuthal}
and {\it tri-azimuthal} cases only. The coordinates are parametrised as follows.

\bigskip
\noindent
\underline{Axial coordinates}:

\medskip
\noindent 
In this case we label the coordinate on $\R^D$ as follows,
\be
\label{axx}
x_{\mu}=(x_i,x_D)\ \ \ ,\quad 
i=1,2,..,D-1\quad,\quad|x_i|^2=r^2\quad,\quad
R^2=r^2+x_D^2\,,
\ee
so that
\be
\label{rt}
r=R\sin\ta_{1}\quad,\quad x_D=R\cos\ta_{1}\,,
\ee
where $\ta_{1}$ is the leading polar angle in each dimension, 
parametrised by
the spherical polar angles 
$(\ta_{1},\ta_{2},...,\ta_{D-3},\ta_{D-2},\vf)$,
$\vf$ being the azimuthal angle (with $0\leq \vf \leq 2 \pi,~0\leq \ta_i\leq\pi$)
Our definition of axial symmetry amounts to spherical symmetry in the 
$D-1$
dimensional subspace, as for example in \cite{Witten:1976ck}.

%%%%%%%%%%%%%%%%%%%%%%%%%%%%%
\bigskip
\noindent
\underline{Azimuthal coordinates}: 

\medskip
\noindent
Imposing azimuthal symmetry in the $x_i=(x_1,x_2)$
plane, leaves the dependence of the fields on the coordinates 
$x_I=(x_3,x_4,..,x_D)$
unrestricted. In practice however, we will restrict to the $D=4$ case 
only
for reasons explained in Section {\bf 4} below. The labeling we will 
employ is
\be
\label{cylx}
x_{\mu}=(x_i,x_I)\ \ \ ,\quad i=1,2\ \ \ ,\quad I=3,4\quad,\quad
|x_i|^2=\rho^2\quad,\quad R^2=\rho^2+|x_I|^2\,,
\ee
so that
\be
\label{rhoI}
\rho=R\sin\ta_{1}\sin\ta_{2}\quad,\quad 
x_3=R\sin\ta_{1}\cos\ta_{2}\quad,\quad
x_4=R\cos\ta_{1}\,,
\ee
or
\be
\label{rhoI1}
\rho=r\,\sin\ta_{2}\quad,\quad x_3=z=r\,\cos\ta_{2}\quad,\quad
x_4=t\,.
\ee

\bigskip
\noindent
\underline{Intermediate coordinates}:

\medskip
\noindent
In $D=5$ the only {\it intermediate} possibility is $n=3$, and we label the coordinate as
\be
\label{int5}
x_{\mu}=(x_i,x_4,x_5)\equiv(x_i,s,t)\ \ \ ,\quad 
i=1,2,3\quad,\quad|x_i|^2=r^2\quad,\quad
R^2=r^2+s^2+t^2\,,
\ee
so that
\be
\label{rt5}
r=R\sin\ta_{1}\sin\ta_{2}\quad,\quad s=R\sin\ta_{1}\cos\ta_{2}\quad,\quad
t=R\cos\ta_1\,,
\ee
in an angular parametrisation $(\ta_{1},\ta_{2},\ta_{3},\vf)$, with polar angles ranging from
$0$ to $\pi$, $\vf$ being the azimuthal angle ranging from $0$ to $2\pi$. The notation
\be
\label{hat5}
\hat x_i=(\sin\ta_3\cos\vf,\sin\ta_3\sin\vf,\cos\ta_3)
\ee
will be employed below.

In $D=6$ both $n=3$ and $n=4$ are possible, but the second leads to a $4$ dimensional effective
system which is superfluous for our purposes here. Hence we restrict to $n=3$ and label the coordinate as
\be
\label{int6}
x_{\mu}=(x_i,x_5,x_6)\equiv(x_i,s,t)\ \ \ ,\quad 
i=1,2,3,4\quad,\quad|x_i|^2=r^2\quad,\quad
R^2=r^2+s^2+t^2\,,
\ee
so that $(r,s,t)$ are parametrised exactly as in \re{rt5}
in an angular parametrisation $(\ta_{1},\ta_{2},\ta_{3},\ta_{4},\vf)$. The notation
\be
\label{hat6}
\hat x_i=(\sin\ta_3\sin\ta_4\cos\vf,\sin\ta_3\sin\ta_4\sin\vf,\sin\ta_3\cos\ta_4,\cos\ta_3)
\ee
being employed for this case below.

\bigskip
\noindent
\underline{Bi-azimuthal coordinates}: 

\medskip
\noindent
In this case we will restrict our attention to $D=4$ and $D=5$. (Bi-azimuthal symmetry in $D=6$ would
lead to four dimensional residual subsystems, which are superfluous for our purposes here.)
In the first case we will subject
the components of the energy density functional \re{blocks} to the symmetry implied by the
Ansatz, while in the second, we will only state the Ansatz since no solutions will be constructed
subsequently in that case.

In $D=4$ we impose a second azimuthal symmetry in \re{cylx}, in the $x_I=(x_3,x_4)$, plane,
denoting the radial variable in the $(x,y)$ and $(z,t)$
planes with $\rho=\sqrt{x^2+y^2}=\sqrt{|x_i|^2}$ and $\si=\sqrt{z^2+t^2}=\sqrt{|x_I|^2}$.
In this case, we will parametrise $\R^4$ as
\bea
x_{i}&=&(R\sin\psi)\,\,\,\hat x_{i}\equiv\rho\,\hat x_{i}\quad,\quad
\hat x_{i}=(\cos\vf_1,\sin\vf_1)
\label{xi}
\\
\nonumber
x_{I}&=&(R\cos\psi)\,\,\,\hat x_{I}\equiv\si\hat x_I\quad,\quad
\hat x_{I}=(\cos\vf_2,\sin\vf_2)
\label{xI}
\eea
where $R^2=|x_i|^2+|x_I|^2=|x_{\mu}|^2$, with 
$0\le\psi\le\frac{\pi}{2}$, $0\le\vf_1\le 2\pi$ and $0\le\vf_2\le 2\pi$. While the two angles $(\vf_1,\vf_2)$
are azimuthal angles, the angle $\ta$ here is not a polar angle as its range is one half of that of a polar
angle. We shall refer to such angles as semi-polar henceforth.

In $D=5$, $\R^5$ is parametrised as
\bea
x_{i}&=&(R\sin\ta\sin\psi)\,\,\,\hat x_{i}\equiv r\,\,\hat x_{i}\quad,\quad
\hat x_{i}=(\cos\vf_1,\sin\vf_1)
\label{xi5}
\\
\nonumber
x_{I}&=&(R\sin\ta\cos\psi)\,\,\,\hat x_{I}\equiv s\,\,\hat x_I\quad,\quad
\hat x_{I}=(\cos\vf_2,\sin\vf_2)
\label{xI5}
\\
x_5&=&R\cos\ta\equiv t
\label{x55}
\eea
where $R^2=r^2+s^2+t^2$, and $0\le\ta\le\pi$ and $0\le\psi\le\frac{\pi}{2}$. In \re{xi5}-\re{x55}, $\ta$
is a polar angle and $\psi$ a semi-polar angle. 
We denote polar angles by $\ta$ and semi-polar angles by
$\psi$ henceforth. All azimuthal angles are likewise denoted by $\vf$.

\bigskip
\noindent
\underline{Tri-azimuthal coordinates}:

\medskip
\noindent
Here we restrict our attention to $D=6$ only for reasons explained already.
Extending the labeling \re{xi}  of $\R^4$ to that of $\R^6$, 
with
$\rho=\sqrt{x_1^2+x_2^2}=\sqrt{|x_{i_1}|^2}$, $i_1=1,2$, 
$\si=\sqrt{x_3^2+x_4^2}=\sqrt{|x_{i_2}|^2}$,
$i_2=3,4$, and $\tau=\sqrt{x_5^2+x_6^2}=\sqrt{|x_{i_3}|^2}$, 
$i_3=5,6$, by
\bea
\nonumber
x_{i_1}&=&(R\sin\psi_1\sin\psi_2)\,\,\,\hat x_{i_1}\equiv\rho\,\hat 
x_{i_1}\quad,\quad
\hat x_{i_1}=(\cos\vf_1,\sin\vf_1)
\label{x1}
\\
x_{i_2}&=&(R\sin\psi_1\cos\psi_2)\,\,\,\hat x_{i_2}\equiv\si\hat 
x_{i_2}\quad,\quad
\hat x_{i_2}=(\cos\vf_2,\sin\vf_2)
\label{x2}
\\
\nonumber
x_{i_3}&=&(R\cos\psi_1)\,\,\,\hat x_{i_3}\,\,\,\qquad\equiv\tau\,\hat 
x_{i_3}\quad,\quad
\hat x_{i_3}=(\cos\vf_3,\sin\vf_3)
\label{x3}
\eea
where $R^2=|x_{i_1}|^2+|x_{i_2}|^2+|x_{i_3}|^2=|x_{\mu}|^2$,
with $0\le\psi_1\le\frac{\pi}{2}$, $0\le\psi_2\le\frac{\pi}{2}$, and 
with the three
azimuthal angles $0\le\vf_1\le 2\pi$, $0\le\vf_2\le 2\pi$ and 
$0\le\vf_3\le 2\pi$.

%%%%%%%%%%%%%%%%%%%%%%%%%%%%%%%%%%%%%%%%%%%%%%%%%%%%%%%%%%%%%%%%%%
\subsection{Spherical symmetry}
%%%%%%%%%%%%%%%%%%%%%%%%%%%%%%%%%%%%%%%%%%%%%%%%%%%%%%%%%%%%%%%%%%
The spherically symmetric Ansatz for the scalar field $\f^a$ in $D$ 
dimensions is
\be
\label{sph}
\f^a=\eta\,Q(R)\,\hat x^a\quad,\quad\hat x^a=\frac{x^a}{R}\,,
\ee
resulting in the reduced building blocks \re{blocks}
\bea
\nonumber
\left|\f_{\mu}^{a}\right|^2&=&Q_R^2+(D-1)\left(\frac{Q}{R}\right)^2
\label{1s}
\\
\left|\f_{\mu\nu}^{ab}\right|^2&=&2(D-1)\left(\frac{Q}{R}\right)^2
\left[2\,Q_R^2+(D-2)\left(\frac{Q}{R}\right)^2\right]
\label{2s}
\\
\nonumber
\left|\f_{\mu\nu\rho}^{abc}\right|^2&=&6(D-1)(D-2)\left(\frac{Q}{R}\right)^4
\left[3\,Q_R^2+(D-3)\left(\frac{Q}{R}\right)^2\right]
\label{3s}
\eea
where we have used the notation $Q_R=\frac{\pa Q}{\pa R}$\,.

%%%%%%%%%%%%%%%%%%%%%%%%%%%%%%%%%%%%%%%%%%%%%%%%%%%%%%%%%
\subsection{Axial symmetry}
The axially symmetric Ansatz for the scalar field 
$\f^a=(\f^{\al},\f^D)$ in $D\ge 4$
dimensions, with the index $\al=1,2,..,D-1$ is
\be
\label{ax}
\f^{\al}=\eta\,H(r,x_D)\,\hat x^{\al}
\quad,\quad\f^D=\eta\,G(r,x_D)\quad,\quad\hat 
x^{\al}=\frac{x^{\al}}{r}\,,
\ee
using the labeling \re{axx} of the coordinates.

There is a very important exception in the $D=3$ case of \re{ax}, 
where imposition of
axial symmetry on the field $\f^a=(\f^A,\f^3)$, $A=1,2$, is 
tantamount to imposing
azimuthal symmetry. The axially symmetric Ansatz in $D=3$ is
\be
\label{ax3}
\f^{A}=\eta\,H(r,x_3)\,n^{A}
\quad,\quad\f^3=\eta\,G(r,x_3)\quad,\quad n^A
=\left[\begin{array}{c}
\cos n\vf \\
\sin n\vf
\end{array}\right]\,,
\ee
$n=1,2,3,...$ being the azimuthal vortex number.

The result of substituting \re{ax} in \re{blocks} is
\bea
%\nonumber
\left|\f_{\mu}^{a}\right|^2&=&\left(H_r^2+G_r^2+H_D^2+G_D^2\right)+
(D-2)\left(\frac{H}{r}\right)^2~,
\label{2a}
\\
\frac12\left|\f_{\mu\nu}^{ab}\right|^2&=&2(H_{[r}\,G_{D]})^2+
(D-2)\left(\frac{H}{r}\right)^2\left[2\left(H_r^2+G_r^2+H_D^2+G_D^2\right)+
(D-3)\left(\frac{H}{r}\right)^2\right]~,
\label{2a1}
\nonumber
\\
\nonumber
\frac16\left|\f_{\mu\nu\rho}^{abc}\right|^2&=&(D-2)
\left(\frac{H}{r}\right)^2\Bigg\{6(H_{[r}\,G_{D]})^2+
\nonumber
\\
\nonumber
\qquad\qquad\qquad&{~~}&+
(D-3)\left(\frac{H}{r}\right)^2\left[3\left(H_r^2+G_r^2+H_D^2+G_D^2\right)+
(D-4)\left(\frac{H}{r}\right)^2\right]\Bigg\}~,
\label{3a}
\eea
where we have used the notation $H_r=\frac{\pa H}{\pa r}$,
$H_D=\frac{\pa H}{\pa x_D}$, and $H_{[r}\,G_{D]}=(H_rG_D-H_DG_r)$.

The spherically symmetric limit  \re{2s} of \re{2a}  
follows
immediately from \re{rt}, by the replacements
\[
H(r,x_D)=Q(R)\,\sin\ta_{D-2}\quad\quad,
G(r,x_D)=Q(R)\,\cos\ta_{D-2}\,,
\]
with $R^2=r^2+x_D^2$, and using
\[
\pa_r=\sin\ta_{D-2}\,\pa_R+\frac{\cos\ta_{D-2}}{R}\,\pa_{\ta_{D-2}}\quad,\quad
\pa_D=\cos\ta_{D-2}\,\pa_R-\frac{\sin\ta_{D-2}}{R}\,\pa_{\ta_{D-2}}\,.
\]

%%%%%%%%%%%%%%%%%%%%%%%%%%%%%%%%%%%%%%%%%%%%%%%%%%%%%%%%%%%%%%%%%%
\subsection{Azimuthal symmetry}
%%%%%%%%%%%%%%%%%%%%%%%%%%%%%%%%%%%%%%%%%%%%%%%%%%%%%%%%%%%%%%%%%%
This subsection is concerned with the imposition of azimuthal symmetry in a $D$ dimensional
system, resulting in a $D-1$ dimensional residual subsystem. As such, it does not lead to a
boundary value problem which can be tackled numerically in a practical way. It should thus be
viewed as a first step towards the imposition of bi-azimuthal symmetry in the $D=2+2=4$ case
presented in the next subsection.

Imposing azimuthal symmetry in the $x_i=(x_1,x_2)$ subspace (plane) 
of
$x_{\mu}=(x_i,x_I)$, $I=3,4,...,D$, and labeling the scalar field as
$\f^a=(\f^A,\f^{A'})$, $A=1,2$ and $A'=3,4,...,D$,
the components $\f^A$ are restricted by the Ansatz
\be
\label{azim}
\f^{A}=h(\rho,x_I)\,n^{A}\quad,\quad
n^{A}=(\cos n\vf,\sin n\vf)\quad,\quad\rho^2=|x_i|^2=x^2+y^2\,,
\ee
while the $D-2$ components $\f^{A'}=\f^{A'}(\rho,x_I)$ retain their 
dependence on the
$D-2$ coordinates $x_I$.

The result of enforcing the Ansatz \re{azim} is most compactly expressed by employing
the coordinate $x_M=(x_I,\rho)$, and by labeling the residual field as $\chi^{\al}=
(\chi^{A'},\chi^{D-1})\equiv(\f^{A'},h)$, with the new index running over $\al=A',D-1$.
In this notation we have
\bea
\nonumber
\left|\f_{\mu}^{a}\right|^2&=&\left(\frac{n\chi^{D-1}}{\rho}\right)^2+
|\pa_M\chi^{\al}|^2
\label{1c}
\\
\frac12\left|\f_{\mu\nu}^{ab}\right|^2&=&4\left(\frac{n\chi^{D-1}}{\rho}\right)^2
|\pa_M\chi^{\al}|^2+|\pa_{[M}\chi^{\al}\pa_{N]}\chi^{\beta}|^2
\label{2c}
\\
\nonumber
\frac16\left|\f_{\mu\nu\rho}^{abc}\right|^2&=&\frac52
\left(\frac{n\chi^{D-1}}{\rho}\right)^2
|\pa_M\chi^{\al}|^2+|\pa_{[M}\chi^{\al}\pa_{N]}\chi^{\beta}|^2
+\frac16|\pa_{[M}\chi^{\al}\pa_{N}\chi^{\beta}\pa_{R]}\chi^{\gamma}|^2
\label{3c}
\eea

In the case of interest here, namely for $D=4$,
$x_{\mu}=(x_i,x_I)$, with $i=1,2=x,y$ and $I=3,4=z,t$,
the azimuthally symmetric Ansatz \re{azim} now becomes
\be
\label{cyl}
\f^{A}=h(\rho,z,t)\,n^{A}\quad,\quad\f^{A'}=\chi^{A'}(x_I,\rho)
=\left[\begin{array}{c}
f(\rho,z,t) \\
g(\rho,z,t)
\end{array}\right]\,,
\ee
resulting in the residual $3$ dimensional system with coordinates 
$x_M=(z,t,\rho)$
being given by \re{2c} with $\chi^{D-1}=\chi^3\equiv h$.

The axially symmetric limit \re{2a}  of  \re{2c} 
follows
immediately from \re{rhoI}, by the replacements
\[
h(z,t,\rho)=H(r,t)\,\sin\ta_{1}\quad ,\quad
f(z,t,\rho)=H(r,t)\,\cos\ta_{1}\quad , \quad g(z,t,\rho)=G(r,t)\,,
\]
with $r^2=\rho^2+z^2$, and using
\[
\pa_{\rho}=\sin\ta_{1}\,\pa_r+\frac{\cos\ta_{1}}{r}\,\pa_{\ta_{1}}\quad,\quad
\pa_z=\cos\ta_{1}\,\pa_r-\frac{\sin\ta_{1}}{r}\,\pa_{\ta_{1}}\,.
\]

%%%%%%%%%%%%%%%%%%%%%%%%%%%%%%%%%%%%%%%%%%%%%%%%%%%%%%%%%%%%%%%%%%
\subsection{Intermediate symmetries}
%%%%%%%%%%%%%%%%%%%%%%%%%%%%%%%%%%%%%%%%%%%%%%%%%%%%%%%%%%%%%%%%%%
The symmetries to be considered here are rotational 
symmetry in the $\R^3$ subspace of $\R^5$ (for $D=5$),
and the $\R^4$ subspace of $\R^6$ (for $D=6$). 
(Rotational symmetry in the $\R^3$ subspace of $\R^6$
would be superfluous since that would lead to a four dimensional effective system.)

For $D=5$, the intermediate symmetric Ansatz for the field $\f^a=(\f^{\al},\f^4,\f^5)$ is
\be
\label{ansatz_int5}
\f^{\al}=\eta\,h(r,s,t)\,\hat x^{\al}\quad,\quad\f^4=\eta\,g(r,s,t)\quad,\quad\f^5=\eta\,f(r,s,t)
\ee
using the notation of \re{int5}-\re{hat5}.

The intermediate symmetric Ansatz for the field $\f^a=(\f^{\al},\f^5,\f^6)$ for $D=6$ is
\be
\label{ansatz_int6}
\f^{\al}=\eta\,h(r,s,t)\,\hat x^{\al}\quad,\quad\f^5=\eta\,g(r,s,t)\quad,\quad\f^6=\eta\,f(r,s,t)
\ee
which looks formally identical to \re{ansatz_int5}, but now the coordinates being read from
\re{int6}-\re{hat6}.

In both cases the system reduces to a three dimensional effective subsystems, for which
numerical constructions are outside the scope of this work. 
Hence we do not present the result of symmetry imposition on
the energy density functionals \re{blocks}.

%%%%%%%%%%%%%%%%%%%%%%%%%%%%%%%%%%%%%%%%%%%%%%%%%%%%%%%%%%%%%%%%%%
\subsection{Bi-azimuthal symmetry}
%%%%%%%%%%%%%%%%%%%%%%%%%%%%%%%%%%%%%%%%%%%%%%%%%%%%%%%%%%%%%%%%%%
Our considerations in this subsection cover two cases, namely to state the bi-azimuthal Ans\"atze in
$D=4$ and $D=5$. The residual subsystem in each case is two dimensional and three dimensional, respectively.
In the first case we will construct the solutions numerically, so the Ansatz will be imposed on the
energy density functional, while in the second we will limit ourselves to stating the Ansatz.

\medskip
\noindent
\underline{Bi-azimuthal symmetry in $D=4$}\,:

In the $D=4$ case, using the notation \re{xi} for the coordinates and using the 
same notation \re{azim} as in subsection {\bf 3.3}, $\f^a=(\f^A,\f^{A'})$, the bi-azimuthally 
symmetric Ansatz is
\bea
\f^A&=&\eta\,h(\rho,\si)\,n_1^A\quad,\quad
n_1^A=(\cos n_1\vf_1,\sin n_1\vf_1)~,
\label{fa}
\\
\nonumber
\f^{A'}&=&\eta\,g(\rho,\si)\,n_2^{A'}\quad,\quad
n_2^{A'}=(\cos n_2\vf_2,\sin n_2\vf_2)\,,
\label{fa'}
\eea
where $n_1$ and $n_2$ are the respective vorticities in the two 
planes.

In fact the Ansatz \re{fa}  results in the first stage from the imposition of
azimuthal symmetry \re{azim} in $D=4$,  with the residual fields $\chi^{\al}=(\f^{A'},h)$,
and then imposing a second stage of azimuthal symmetry on the triplet $\chi^{\al}$.
Concerning the imposition of the second stage of azimuthal symmetry, we point out that
the densities \re{2c}  resulting from the first stage do not exhibit a global $SO(D-1)$
invariance, although the original densities \re{blocks} are invariant under a global
$SO(D)$~\footnote{This is in contrast to that of a YM system, where the local gauge group does,
under azimuthal symmetry imposition, reduce to an effective YM-Higgs
system exhibiting a broken local gauge invariance~\cite{Radu:2006gg}.}. We have verified that
the second stage results a consistent reduction, even though the reduced system after the first
stage did not possess a global invariance.

Imposition of bi-azimuthal symmetry enables a $2$ dimensional boundary value problem, to be tackled
numerically in the next section, so we list the resulting densities \re{blocks}
\bea
%\nonumber
\left|\f_{\mu}^{a}\right|^2&=&\left[\left(\frac{n_1h}{\rho}\right)^2+
\left(\frac{n_2g}{\si}\right)^2\right]+
\left(h_{\rho}^2+g_{\rho}^2+h_{\si}^2+g_{\si}^2\right)~,
\label{2b}
\\
\nonumber
\frac{1}{2!^2}\left|\f_{\mu\nu}^{ab}\right|^2&=&
\left(\frac{n_1h}{\rho}\right)^2\left(\frac{n_2g}{\si}\right)^2+
\left[\left(\frac{n_1h}{\rho}\right)^2+\left(\frac{n_2g}{\si}\right)^2\right]
\left(h_{\rho}^2+g_{\rho}^2+h_{\si}^2+g_{\si}^2\right)+
\left(h_{[\rho}\,g_{\si]}\right)^2~,
\label{1b}
\\
\nonumber
\frac{1}{3!^2}\left|\f_{\mu\nu\rho}^{abc}\right|^2&=&
\left(\frac{n_1h}{\rho}\right)^2\left(\frac{n_2g}{\si}\right)^2
\left(h_{\rho}^2+g_{\rho}^2+h_{\si}^2+g_{\si}^2\right)+
\left[\left(\frac{n_1h}{\rho}\right)^2+\left(\frac{n_2g}{\si}\right)^2\right]
\left(h_{[\rho}\,g_{\si]}\right)^2~,
\label{3b}
\eea
where we have used the notation $h_{\rho}=\frac{\pa h}{\pa\rho}$,
$h_{\si}=\frac{\pa h}{\pa\si}$, and 
$g_{[\rho}\,g_{\si]}=(h_{\rho}g_{\si}-g_{\rho}h_{\si})$
as in \re{2a}.

In terms of the coordinates $\rho=R\sin\ta$, $\si=R\cos\ta$ defined 
by \re{xi}, the spherically symmetric limit \re{2a} of \re{2b} follows
immediately from by the replacements
\[
h(\rho,\si)=Q(R)\sin\ta\quad,\quad g(\rho,\si)=Q(R)\cos\ta\,,
\]
and using
\[
\pa_{\rho}=\sin\ta\,\pa_R+\frac{\cos\ta}{R}\,\pa_{\ta}\quad,\quad
\pa_{\si}=\cos\ta\,\pa_R-\frac{\sin\ta}{R}\,\pa_{\ta}\,.
\]
This limit will be exploited in the numerical constructions.

\medskip
\noindent
\underline{Bi-azimuthal symmetry in $D=5$}\,:

Here, the residual system being three dimensional, we simply state the Ansatz
\bea
\f^A&=&\eta\,h(r,s,t)\,n_1^A\quad,\quad
n_1^A=(\cos n_1\vf_1,\sin n_1\vf_1)~,
\\
\f^{A'}&=&\eta\,g(r,s,t)\,n_2^{A'}\quad,\quad
n_2^{A'}=(\cos n_2\vf_2,\sin n_2\vf_2)~,
\label{fa5}
\\
\f^5&=&\eta\,f(r,s,t)\,,
\eea
in the notation of \re{xi5}-\re{x55}.

%%%%%%%%%%%%%%%%%%%%%%%%%%%%%%%%%%%%%%%%%%%%%%%%%%%%%%%%%%%%%%%%%%
\subsection{Tri-azimuthal symmetry}
%%%%%%%%%%%%%%%%%%%%%%%%%%%%%%%%%%%%%%%%%%%%%%%%%%%%%%%%%%%%%%%%%%
As noted at the start of this section, we shall simply state the 
Ansatz here, for $6$ dimensions only,
without imposing the symmetry on the energy density building blocks 
\re{blocks}. Then in the next
section we will use this to calculate the topological charge of the 
putative solutions in $6$ dimensions,
which are not constructed numerically here. The tri-azimuthally 
symmetric Ansatz for
$\f^a=(\f^{A_1},\f^{A_2},\f^{A_3})$, $A_1=1,2$, $A_2=3,4$, 
$A_3=5,6$
\bea
\nonumber
\f^{A_1}&=&h(\rho,\si,\tau)\,n_1^{A_1}\quad,\quad
n_1^{A_1}=(\cos n_1\vf_1,\sin n_1\vf_1)
\label{fa1}
\\
\f^{A_2}&=&g(\rho,\si,\tau)\,n_2^{A_2}\quad,\quad
n_2^{A_2}=(\cos n_2\vf_2,\sin n_2\vf_2)
\label{fa2}
\\
\nonumber
\f^{A_3}&=&f(\rho,\si,\tau)\,n_2^{A_3}\quad,\quad
n_3^{A_3}=(\cos n_3\vf_3,\sin n_3\vf_3)
\label{fa3}
\eea
where $n_1$, $n_2$ and $n_3$ are the respective vorticities in the 
three planes $(x_1,x_2)$,
$(x_3,x_4)$ and $(x_5,x_6)$.

%%%%%%%%%%%%%%%%%%%%%%%%%%%%%%%%%%%%%%%%%%%%%%%%%%%%%%%%%%%%%%%%%%
\section{Topological charges and boundary values}
%%%%%%%%%%%%%%%%%%%%%%%%%%%%%%%%%%%%%%%%%%%%%%%%%%%%%%%%%%%%%%%%%%
In this Section, we present in detail the topological charges resulting from the
various types of boundary values of the scalar field. This is relevant when subjecting
the fields to axial, azimuthal intermediate bi-azimuthal 
and tri-azimuthal symmetries in turn.
Under each (symmetry) heading, we will calculate the topological 
charges in all dimensions $D$ for which
the residual subsystem is at most {\bf three} dimensional. 
This will cover the generic cases, all further
examples being superfluous. Subject to axial symmetry, 
we consider the cases $D=3,4,5,6$. Subject to azimuthal
symmetry, we cover only $D=4$. 
Subject to intermediate symmetry, we take the cases $D=5,6$. 
For configurations with bi-azimuthal
symmetry, we cover $D=4,5$. 
Subject to tri-azimuthal symmetry, we cover the only possible case $D=6$.

As explained at the end of Section {\bf 2}, it is sufficient to calculate the winding
numbers since the topological charges are simply numerical multiples of the latter. Up
to angular volume normalisations $N_D$, these are the surface integrals of the currents
\re{wnf}, hence what we need to calculate are the asymptotic values of the quantities
$\hat x_{\mu}\,\omega^{(D)}_{\mu}$ to enable us to evaluate the surface integrals
\be
\label{surf}
I_D=\int\hat x_{\mu}\,\omega^{(D)}_{\mu}\bigg|_{R=\infty}\,R^{D-1}\,
d\Omega(\ta_{D-2},\ta_{D-3},...,\ta_{1},\vf)\,,
\ee
$\hat x_{\mu}$ being the unit vector, and
$d\Omega(\ta_{D-2},\ta_{D-3},...,\ta_{1},\vf)$ the angular volume 
element, in $\R^D$.

Here, we will evaluate the angular integrals \re{surf}, ($a$) subject 
to
axial symmetry for $D=3,4,5,6$, ($b$) subject to azimuthal symmetry 
for $D=4$, and ($c$)
subject to bi-azimuthal symmetry for $D=4$.

\subsection{Axial symmetry}
In the case of axially symmetric fields, we will impose the following 
asymptotic
boundary values on the functions $H(r,x_D)$ and $G(r,x_D)$ defined in 
\re{ax} for
$D\ge 4$, and in \re{ax3} for $D=3$
\bea
\label{axbv}
\lim_{R\to\infty}H(r,x_D)&=&\sin m\,\ta_{1}
\label{axbvH}
\\
\nonumber
\lim_{R\to\infty}G(r,x_D)&=&\cos m\,\ta_{1}
\label{axbvG}
\quad,\quad m=1,2,3,...
\eea
The topological charges $q_D$ of the axially symmetric models in $D=3,4,5$ are defined by the
integrals \re{surf}, divided by the angular volumes $\Omega_{D-1}=2\pi,2\pi^2,\frac{8\pi^2}{3}$
in each of these dimensions respectively, by the (volume) integrals
\[
q_D=\frac{I_D}{\Omega_{D-1}}=D!\,\int 
H^{D-2}\left(G_RH_{\ta_1}-H_RG_{\ta_1}\right)\,dR\,d\ta_1~,
\] 

The surface integrals \re{surf} can be evaluated analytically. In the 
axially symmetric
cases at hand, where the corresponding volume integrals are two 
dimensional, these become
contour integrals in the positive half plane $r[0,\infty)$, 
$x_D(-\infty,+\infty)$
by virtue of Stokes' theorem. Now the line integral along the $x_D$ 
axis does not contribute
since analiticity requires that $H(\ta_1=0)=H(\ta_1=\pi)=0$, so the 
only contribution comes
from the infinite semicircle, thus reducing \re{surf} to the 
following one dimensional angular
integrals
\be
\label{ang}
I_D^{\rm{axial}}=D!\,\Omega_D\,
\int\,H^{D-2}\,\left(G\,H_{\ta_1}-H\,G_{\ta_1}\right)\big|_{R=\infty}\,d\ta_1\,,
\ee
with the exception of the $D=3$ case where axial symmetry coincides 
with azimuthal
symmetry, when
\be
\label{ang3}
I_3=2!\,2\pi\,n\,\int\,H\,\left(G\,H_{\ta}-H\,G_{\ta}\right)\big|_{R=\infty}\,d\ta\,.
\ee
Subject to the axially symmetric boundary values 
\re{axbvH}, the integrals
\re{ang3} and \re{ang} for $D=3$ and $D=4,5,6$ are evaluated as
\bea
I_3^{\rm{ax}}&=&4\,\eta^3\,\pi\,n\,[1-(-1)^m]
\label{i3ax}
\\
I_4^{\rm{ax}}&=&12\,\eta^4\,\pi^2\,m
\label{i4ax}
\\
I_5^{\rm{ax}}&=&32\,\eta^5\,\pi^2\,[1-(-1)^m]
\label{i5ax}
\\
I_6^{\rm{ax}}&=&5!\,\eta^6\,\pi^3\,m
\label{i6ax}\,.
\eea

We now see from \re{i3ax} and \re{i5ax}, that in odd $D$ dimensions axially symmetric
fields are capable of supporting {\it zero topological charge} solutions describing
an even number $m$ of soliton--antisoliton energy/charge concentrations,
as well as {\it unit topological charge} solutions describing chains~\cite{KKS}
for an odd number $m$. As we shall see from the numerical work in Section {\bf 5.2}, these
concentrations are located slightly off the $x_D$ axis, forming rings analogous to
the nodes on the symmetry axis~\footnote{Subsequent to the construction of zro
charge monopole--antimonopole pairs~\cite{R,KK}, such charge chains of monopoles and antimonopoles
of {\it unit} topological charge were contructed in \cite{KKS}.} found in the $3$ dimensional
Yang-Mills--Higgs case.
We see by contrast from \re{i4ax} and \re{i6ax}, that in even $D$ dimensions axially
symmetric fields are {\bf not} capable of supporting {\it zero topological charge}
solutions. They describe only multisoliton solutions of topological charges $m$,
the concentrations of charge/energy being located on the $x_D$ axis. (These are the
analogues of Witten's axially symmetric instantons~\cite{Witten:1976ck}.) Our numerical solutions in
the next Section will bear out these conclusions.

Having described candidates for {\it zero topological charge} solutions in odd dimensions,
we proceed to explore prescriptions whereby such solutions in even $D$ dimensions can
also be constructed. This is possible only if less stringent symmetry than axial symmetry
is imposed on the system, and below we describe two such distinct prescriptions in $D=4$,
employing in turn azimuthal and bi-azimuthal symmetries, and one such prescription in $D=6$
employing tri-azimuthal symmetry.

\subsection{Azimuthal symmetry}
In the case of azimuthal symmetry in $D=4$, the asymptotic boundary
values to be imposed on the functions $h(\rho,z,t)$, $f(\rho,z,t)$ 
and $g(\rho,z,t)$
defined in \re{cyl} are
\bea
\nonumber
\lim_{R\to\infty}h(\rho,z,t)&=&\sin m_1\ta_1\,\sin m_2\ta_2
\label{cylbvh}
\\
\lim_{R\to\infty}f(\rho,z,t)&=&\sin m_1\ta_1\,\cos m_2\ta_2
\label{cylbvf}
\\
\nonumber
\lim_{R\to\infty}g(\rho,z,t)&=&\cos m_1\ta_1
\label{cylbvg}\quad,\quad m_1,m_2=1,2,3,...
\eea

We note here that the asymptotic axially symmetric boundary values 
are described by
one integer $m$ for $D\ge 4$ and two integers $(m,n)$ for $D=3$, 
while those for the
azimuthal boundary values for $D=4$ are given in terms of the triple 
of integers
$(m_1,m_2,n)$.

Substituting azimuthal Ansatz \re{azim} in \re{surf} for $D=4$, and 
using the analyticity
requirement that $h(r,t)$ vanishes on the $t$--axis, this reduces to 
the two dimensional
angular integral
\be
\label{anga}
I_4^{\rm{az}}=4!\,n\,\int 
h\left[h\left(f_{\ta_2}g_{\ta_1}-f_{\ta_1}g_{\ta_2}\right)+
f\left(g_{\ta_2}h_{\ta_1}-h_{\ta_1}h_{\ta_2}\right)+
g\left(h_{\ta_2}f_{\ta_1}-g_{\ta_1}f_{\ta_2}\right)\right]\,d\ta_1\,d\ta_2\,,
\ee
which can readily be evaluated subject to the boundary conditions 
\re{cylbvf}  to
yield
\be
\label{i4cyl}
I_4^{\rm{azim}}=12\,\eta^4\,\pi^2\,m_1\,n\,\left[1-(-1)^{m_2}\right]\,.
\ee
This accommodates both multi-soliton (for {\it odd} $m_2$) and zero 
topological charge (for {\it even}
$m_2$) solutions, labeled by the triple of integers $(m_1,m_2,n)$. 
Unfortunately the numerical solution
the corresponding field equations involves three dimensional 
integration, which task is beyond
the scope of the present work.

\subsection{Intermediate symmetries}

In both the $D=5$ and $D=6$ cases discussed in section {\bf 3.4} above, the asymptotic behaviours
consistent with finite energy are both stated formally as
\bea
\nonumber
\lim_{R\to\infty}h(r,s,t)&=&\sin m_1\ta_1\,\sin 
m_2\ta_2
\label{intbvh}
\\
\lim_{R\to\infty}g(r,s,t)&=&\sin m_1\ta_1\,\cos 
m_2\ta_2
\label{intbvg}
\\
\nonumber
\lim_{R\to\infty}f(r,s,t)&=&\cos m_1\ta_1
\label{intbvf}\quad,\quad 
m_1,m_2=1,2,3,...
\eea
augmented by the analyticity condition $h(r=0)=0$, which is crucial in the evaluation of the surface
intergals. In both the $D=5$ and $D=6$ here, these follow from three dimensional volume integrals which
are formally identical. Up to numerical factors, these are expressed as
\bea
\nonumber
I_{5,6}^{\rm{inter}}&\sim&\eta^{5,6}\,(\pi)^{2,3}\,m_1\,m_2\,\int\vep_{\mu\nu\rho}\,\vep^{ABC}\,
\pa_{\mu}\Xi^A\,\pa_{\nu}\Xi^B\,\pa_{\rho}\Xi^C\,dr\,ds\,dt
\label{intvol}
\\
&=&\eta^{5,6}(\pi)^{2,3}\,m_1\,m_2\,\int
\vep_{\mu\nu\rho}\vep^{ABC}\,\Xi^A\,\pa_{\nu}\Xi^B\,\pa_{\rho}\Xi^C\,dS_{\mu}~,
\label{surf56}
\eea
where we have used the notation $x_{\mu}=(r,s,t)$, and, the triplet function
$\Xi^A$, $A=1,2,3$ in the two cases is defined in terms of the functions $(h,g,f)$ as
\[
\Xi^A=\left((h)^3,g,f\right)\quad{\rm and}\quad\Xi^A=\left((h)^4,g,f\right)\,,
\]
respectively. The nonvanishing contributions to the surface integral(s) \re{surf56} come from the upper
hemisphere.

The values of the respective surface integrals in $D=5,6$ are calculated to be
\bea
I_5^{\rm{inter}}&=&\frac{5\cdot 2^6}{3^2}\,\eta^5\,\pi^2\,m_2\,\left[1-(-1)^{m_1}\right]
\label{i5int}\\
I_6^{\rm{inter}}&=&15\,\eta^6\,\pi^3\,m_1\,\left[1-(-1)^{m_2}\right]\,,
\label{i6int}
\eea
Note the roles of $m_1$ and $m_2$ interchanging in \re{i5int} and \re{i6int}, following from
cancellations ocurring when evaluating \re{intvol}.

One sees again that by relaxing axial symmetry and imposing a weaker symmetry,
it is possible to support both
{\it multisolitons} of arbitrary topological charges, and, {\it soliton--antisolitons} chains,
with {\it zero} and {\it nonzero} topological charges in {\it all dimensions}.
Unfortunately the simplest such examples result in three dimensional boundary value
problems, which is at present technically too hard a task to perform. The situation is the same in the
azimuthal case in $D=4$ above.

\subsection{Bi-azimuthal symmetry}

Bi-azimuthal symmetry will be applied in $D=4$ and $D=5$, each resulting in a two and a three
dimensional subsystems, respectively.

\medskip
\noindent
\underline{$D=4=2+2$}

In this case the fields are described by the bi-azimuthal Ansatz 
\re{fa}. The asymptotic behaviours of the functions $h$ and $g$ in \re{fa} are taken to be
\bea
\label{mh}
\lim_{R\to\infty}\,h=\sin m\psi~,~~~
\lim_{R\to\infty}\,g=\cos m\psi~.
\eea
The topological charge in this case is
\bea
\nonumber
I_4^{\rm{bi-azim}}&=&\eta^4\,3!\,(2\pi)^2\,n_1\,n_2\int\vep_{\mu\nu}\,\vep^{AB}\,
\pa_{\mu}\Xi^A\,\pa_{\nu}\Xi^B\,d\rho\,d\si
\label{volb}
\\
&=&\eta^4\,3!\,(2\pi)^2\,n_1\,n_2\int
\left(\vep^{AB}\,\Xi^A\,\pa_{\mu}\Xi^B\right)ds_{\mu}
\label{line}
\eea
where we have used the notation $x_{\mu}=(\rho,\si)$, and 
$\Xi^A=\left((h)^2,(g)^2\right)$.

Using the analyticity conditions $h(\psi=0)=0$ and $g(\psi=\frac{\pi}{2})=0$ leads to the
vanishing of the line integrals on the $\rho$ and the $\si$ axes, the 
non vanishing
contribution coming from the infinite quarter circle contour, readily 
evaluated
to yield
\be
\label{i4bi}
I_4^{\rm{bi-azim}}=\eta^4\,2\,\pi^2\,n_1n_2\,\left[1-(-1)^m\right]\,,
\ee
which supports both multisolitons and zero charge soliton-antisolitons.

%\medskip
%\newpage
\noindent
\underline{$D=5=2+2+1$}

In this case the fields are described by the bi-azimuthal Ansatz 
\re{fa5}. The asymptotic behaviours of the functions $h$ and $g$ and $f$ in \re{fa5} are taken to be
\bea
\label{mh2}
\lim_{R\to\infty}\,h=\sin m_1\ta\sin m_2\psi~,~~~
\lim_{R\to\infty}\,g=\sin m_1\ta\cos m_2\psi~,~~~
\lim_{R\to\infty}\,f=\cos m_1\ta~.
\eea
The topological charge now reduces to a three dimensional integral in the residual coordinates
$x_{\mu}=(r,s,t)$
\bea
\nonumber
I_5^{\rm{bi-azim}}&=&\eta^5\,5\,(2\pi)^2\,n_1\,n_2\,\int\vep_{\mu\nu\rho}\,\vep^{ABC}\,
\pa_{\mu}\Xi^A\,\pa_{\nu}\Xi^B\,\pa_{\rho}\Xi^C\,dr\,ds\,dt
\label{volt}
\\
&=&\eta^5\,5\,(2\pi)^2\,n_1\,n_2\,\int
\vep_{\mu\nu\rho}\vep^{ABC}\,\Xi^A\,\pa_{\nu}\Xi^B\,\pa_{\rho}\Xi^C\,dS_{\mu}~,
\label{surf5}
\eea
in which the triplet function $\Xi^A$, $A=1,2,3$ is defined as
\[
\Xi^A=((h)^2,(g)^2,f)\,.
\]
The surface integral \re{surf5} is then performed to yield
\be
\label{i5bi}
I_5^{\rm{bi-azim}}=\eta^5\,4!\,n_1\,n_2\,\left[1-(-1)^{m_1}\right]\,,
\ee
describing both multisolitons and soliton antisolitons. Note that only $m_1$, and not $m_2$,
features in \re{i5bi}, due to a cancellation in evaluating \re{volt}.

\subsection{Tri-azimuthal symmetry}
This pertains to $D=6$ only. The asymptotic behaviours of the functions $h$, $g$ and $f$ in the
Ansatz \re{fa2}  are taken to be
\bea
\nonumber
\lim_{R\to\infty}h&=&\sin m_1\psi_1\sin m_2\psi_2
\label{ash}
\\
\lim_{R\to\infty}g&=&\sin m_1\psi_1\cos m_2\psi_2
\label{asg}
\\
\nonumber
\lim_{R\to\infty}f&=&\cos m_1\psi_1
\label{asf}\,.
\eea
The topological charge integral in this case is
\bea
\nonumber
I_6^{\rm{tri-azim}}&=&\eta^6\,90\,(2\pi)^3\,n_1\,n_2\,n_3\int\vep_{\mu\nu\rho}\,\vep^{ABC}\,
\pa_{\mu}\Xi^A\,\pa_{\nu}\Xi^B\,\pa_{\rho}\Xi^C\,d\rho\,d\si\,d\tau
\label{volt1}
\\
&=&\eta^6\,90\,(2\pi)^3\,n_1\,n_2\,n_3\int
\vep_{\mu\nu\rho}\vep^{ABC}\,\Xi^A\,\pa_{\nu}\Xi^B\,\pa_{\rho}\Xi^C\,dS_{\mu}~,
\label{surf6}
\eea
where we have used the notation $x_{\mu}=(\rho,\si,\tau)$, and 
$\Xi^A=\left((h)^2,(g)^2,(f)^2\right)$.

To evaluate the surface integral \re{surf6} we need analytic information which comes from finite energy
conditions. 
While we are not displaying here the energy density functional in terms of the functions
$(h,g,f)$, it is nontheless easy to deduce that $h(\psi_1=0,\psi_2=0)=0$. 
$g(\psi_1=0,\psi_2=\frac{\pi}{2}$ and 
$f(\psi_1=\frac{\pi}{2})=0)=0$. 
These, together with continuity conditions, imply that the flux \re{surf6} 
out of the three quarter planes $(\rho,\si)\ ,\ (\si,\tau)$ and $(\tau,\rho)$ vanishes, and hence the 
only contribution comes from the surface bounding the octant of the two--sphere with radius 
$R=\sqrt{\rho^2+\si^2+\tau^2}$.

Applying the boundary functions \re{asg}  on the asymptotic 
octant the flux \re{surf6} yields
\be
\label{i6tri}
I_6^{\rm{tri-azim}}=\frac{5!\,\pi^3}{2}\ n_1\,n_2\,n_3\ 
\left(\frac12\left[1-(-1)^{m_1}\right]\right)^4
\ \left(\frac12\left[1-(-1)^{m_2}\right]\right)^2\,,
\ee
analogous to \re{i4bi}, like which the topological charge vanishes when either $m_1$ or $m_2$ is
{\it even}, and otherwise it is given by the product of the vortex numbers pertaining to each of the
azimuthal symmetries imposed.

%%%%%%%%%%%%%%%%%%%%%%%%%%%%%%%%%%%%%%%%%%%%%%%%%%%%%%%%%%%%%%%%%%%%%%%%%
\section{Numerical constructions}
%%%%%%%%%%%%%%%%%%%%%%%%%%%%%%%%%%%%%%%%%%%%%%%%%%%%%%%%%%%%%%%%%%%%%%%%%

In this Section we give numerical evidence for the existence spherically symmetric, and
axially symmetric solutions in $D=4,~5$.  In addition we have constructed solutions with bi-azimuthal 
symmetry in $D=4$. The solutions of the corresponding $D=3$ model were presented in 
\cite{Paturyan:2005ik} to which we refer for the latter. 

Of course, the most interesting solutions from the viewpoint of 
understanding zero topological
charge, are the axially symmetric ones, but the spherically symmetric 
ones are also presented mainly
because the equations of motion in that case allow a thorough 
asymptotic analysis underpinning the
numerical work. Also the spherically symmetric solutions   
present useful starting profiles for
the $D=4$ bi-azimuthally symmetric multi-solitons.

%%%%%%%%%%%%%%%%%%%%%%%%
\begin{figure}[h!]
\parbox{\textwidth}
{\centerline{
\mbox{
\epsfysize=10.0cm
\includegraphics[width=92mm,angle=0,keepaspectratio]{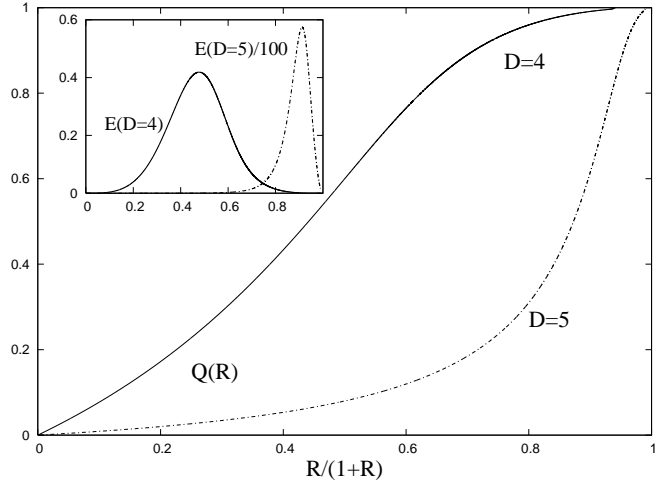} 
}}}
\caption{{\small The  scalar function $Q$ and weighted energy density $E$ of two
typical $D=4$ and $D=5$ spherically
symmetric solutions with $\lambda_1=1$ are shown as a function of the 
compactified radial
coordinate $R/(1+R)$.}}
\end{figure}
%%%%%%%%%%%%%%%%%%%%%%%%
%
Technically, we have restricted ourselves to two dimensional numerical 
integration, solutions with azimuthal
symmetry in $D\ge 4$ representing a difficult numerical challenge 
which we leave for future work.  
Also, one should notice that only one coupling constant $\lambda_i$ 
is relevant here. For example, one may factorize
$\lambda_1$ and, by using a suitable rescaling, one may set 
$\lambda_2=1$ or $\lambda_3=1$ without any loss of generality.

To simplify the picture, 
in this section we shall note $\theta_{D-2}=\theta$ and $x_D=z$.
Also, for all configurations,
the total mass/energy $M$ (which equals the total action)
is computed by integrating the corresponding reduced 
 energy functionals.
%%%%%%%%%%%%%%%%%%%%%%%%%%%%%%%%%%%%%%%%%%%%%%%%%%%%%%%%%%%%%%%%%%%%%%%%%
\subsection{Spherically symmetric solutions}
%%%%%%%%%%%%%%%%%%%%%%%%%%%%%%%%%%%%%%%%%%%%%%%%%%%%%%%%%%%%%%%%%%%%%%%%%
Considering the ansatz (\ref{sph}), the 
reduced
one dimensional weighted energy density reads
\begin{eqnarray}
\label{eredfunc}
 E=R^{D-1}{\cal E}=R^{D-1}\bigg(\lambda_1 
(Q^2-1)^4\left(Q'^2+(D-1)\frac{Q^2}{R^2}\right)
+2(D-2)\lambda_2(Q^2-1)^2  \frac{Q^2}{R^2} 
\\
\nonumber
 \times
\big(2Q'^2+(D-2)\frac{Q^2}{R^2}\big)+6\lambda_3(D-1)(D-2) \frac{Q^4}{R^4} \left(3Q'^2+(D-3) 
\frac{Q^2}{R^2} \right)
\bigg)
\end{eqnarray}
which leads to the following differential equation
\begin{eqnarray}
\nonumber 
& \bigg[2R^{D-1}Q'\left(\lambda_1(Q^2-1)^4+4\lambda_2(D-2)(Q^2-1)^2\frac{Q^2}{R^2}
+18\lambda_3(D-1)(D-2)\frac{Q^4}{R^4}\right) \bigg]'
\\
\nonumber
&=R^{D-2}(2\lambda_1(Q^2-1)^3(4Q'^2+5(D-1)\frac{Q^2}{R^2} -\frac{(D-1)}{R^2}
+4\lambda_2\frac{(D-1)}{R^2}Q(Q^2-1)
\\
\nonumber
&
+36\lambda_3\frac{(D-1)(D-2)}{R^4}Q^3(2Q'^2+(D-3)\frac{Q^2}{R^2}).
\end{eqnarray} 
The solutions of this equation have been constructed 
numerically, for a range of the parameters $\lambda_i$.
We follow the usual approach and, by using a standard ordinary
differential equation solver, we evaluate the initial condition 
\begin{eqnarray}
&&Q(R)=bR-\frac{2b^3\lambda_1}{3(\lambda_1+12b^2(\lambda_2+9b^2\lambda_3))}R^3+O(R^5),
~~{\rm for~}D=4,
\\
&&Q(R)=bR-\frac{2(b^3\lambda_1+4b^5\lambda_2-2)}
{7(\lambda_1+16b^2 \lambda_2+216b^4\lambda_3)}R^3+O(R^5),
~~{\rm for~}D=5
\end{eqnarray} 
 at
$R=10^{-6}$ for global tolerance $10^{-14}$, adjusting for the
shooting
parameter $b$ and integrating  towards  $R\to\infty$.
The behaviour of finite energy solutions for large values of $R$ is
%%%%%%%%%%%%%%%%%%%%%%
\begin{figure}[h!]
\parbox{\textwidth}
{\centerline{
\mbox{
\epsfysize=10.0cm
\includegraphics[width=92mm,angle=0,keepaspectratio]{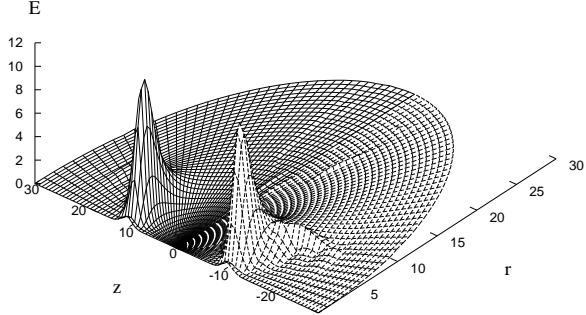} 
}}}
\caption{{\small A three-dimensional plot of the 
weighted energy density 
$E(R,\theta)$ of a $D=5,~m=2$ axially symmetric solution
with $\lambda_1=\lambda_2=1$, $\lambda_3=75$.}}
\end{figure}
%%%%%%%%%%%%%%%%%%%%%%%%
%
%
\begin{eqnarray}
&&Q(R) = 1+ce^{-\frac{2}{3}\sqrt{ {\lambda_2}/{\lambda_3}}R}
-\frac{9\lambda_3}{4\lambda_2}\frac{1}{R^2}
-\frac{243\lambda_3^2 }{16 \lambda_2^2}\frac{1}{R^4} +O(1/R^6),~{\rm 
for~}D=4,\\
&& Q(R) = 1+ce^{-\frac{2}{3}\sqrt{ {\lambda_2}/{\lambda_3}}R}
-\frac{9\lambda_3}{2\lambda_2}\frac{1}{R^2}
+\frac{81\lambda_3^2 }{2 \lambda_2^2}\frac{1}{R^4} +O(1/R^6),~{\rm 
for~}D=5, 
\end{eqnarray} 
where $c$ is a free parameter (the corresponding expressions for the
$D=3$ model are given
in \cite{Paturyan:2005ik}).
For all cases considered, solutions with the correct asymptotics 
are found when the first derivative of the scalar function $Q(R)$ 
evaluated
at the origin, $Q'(0)= b$,  takes on a certain value,
which is a function of $\lambda_i$.  

The profiles of typical $D=4,~5$ solutions are presented in Figure 1 
for $\lambda_1=\lambda_2=\lambda_3=1$. The weighted energy density,
as given by (\ref{eredfunc}) is also exhibited (one should notice the 
different
length scales of the $D=4$ and $D=5$ solitons).
Similar to the 
$D=3$ case, no multinode radial solutions were found, although we 
have no  
analytical argument for their absence.

%%%%%%%%%%%%%%%%%%%%%%%%%%%%%%%%%%%%%%%%%%%%%%%%%%%%%%%%%%%%%%%%%%%%%%%%%
\subsection{Axially symmetric solutions}
%%%%%%%%%%%%%%%%%%%%%%%%%%%%%%%%%%%%%%%%%%%%%%%%%%%%%%%%%%%%%%%%%%%%%%%%
Scalar solitons with axial symmetry are found by taking $m\geq 2$ in 
the
boundary conditions at infinity (\ref{axbvH}).
The two-dimensional weighted energy density $E(R,\theta)=
R^{D-1}\sin^{D-2}\theta~{\cal E}(R,\theta)$ and the
set of two coupled non-linear elliptic partial differential
equations satisfied by the functions $H(R,\theta)$, $G(R,\theta)$ can 
easily be derived by using the 
reduced building blocks (\ref{2a})  and we shall not 
present them here.
These equations are solved numerically, subject to the boundary 
conditions  
\be
\label{r0SA}
H|_{R=0}=0\quad,\quad\pa_{R}G|_{R=0}=0.
\ee 
at the origin and
(\ref{axbvH}) at infinity\footnote{In the numerical algorithm we have
employed a compactified radial coordinate $x=R/(1+R)$,
such that spatial infinity corresponds to $x=1$.}
(we have restricted our analysis to $m=2$ solutions;
 the $m=1$ case corresponds to spherically symmetric 
configurations).
%
%%%%%%%%%%%%%%%%%%%%%%%%
\begin{figure}[h!]
\parbox{\textwidth}
{\centerline{
\mbox{
\epsfysize=10.0cm
\includegraphics[width=92mm,angle=0,keepaspectratio]{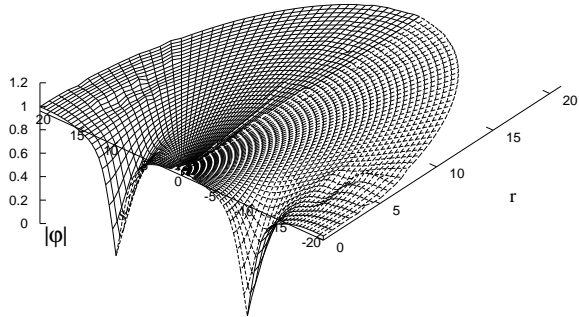} 
}}}
\caption{{\small The modulus of the scalar field 
$|\varphi|=\sqrt{H^2+G^2}$ is shown 
 as a function of the coordinates $r$ and $z$
 for a typical $D=4,~m=2$ axially symmetric solution.
Here $r=R \sin \theta,~z=R\cos \theta$.}}
\end{figure}
%%%%%%%%%%%%%%%%%%%%%%%%%
%
Considering solutions with parity reflection symmetry,
the equations are integrated in the $0\leq \theta \leq \pi/2$ 
region.
The boundary conditions satisfied at the limits of the 
$\theta$-interval are
\begin{eqnarray}
\label{sw1}
H|_{\theta=0}=\partial_{\theta}G|_{\theta=0}=0,
~~\partial_{\theta}H|_{\theta=\pi/2}=G|_{\theta=\pi/2}=0.
\end{eqnarray}
The absence of suitable starting profiles makes this problem extremely 
difficult\footnote{We managed to overcome this difficulty by improving, in succesive steps, an
initial guess solution constructed with suitable trial functions 
which interpolates between the assymptotics (\ref{r0SA}), (\ref{axbvH}).}. 
The numerical calculations were performed with the software package CADSOL/FIDISOL, 
based on the Newton-Raphson method \cite{FIDISOL}.

The  numerical error for the functions is estimated to be 
of the order of $10^{-2}$ or lower for most of the axially symmetric configurations.

Solutions with $m=2$ of the corresponding $D=3$ model were discussed in \cite{Paturyan:2005ik}.
In that case it was possible to distinguish two individual components
(e.g. the modulus of the scalar field $|\phi|=\sqrt{\phi_1^2+\phi_2^2}$
possesses always two distinct zeros on the $z$ axis).

Our $D=4,~5$ results indicate that this is a generic feature of all 
axially symmetric solutions. In Figure 2 we present a three dimensional plot of
the weighted energy density (the reduced Lagrangian) $E(R,\theta)$ of a typical $D=5$, $m=2$ axially
symmetric solution as a function of the $r, z$ (here $\lambda=\lambda_2=1,~\lambda_3=75$).
The modulus of the scalar field $|\varphi|=\sqrt{H^2+G^2}$ of a $D=4$ solution with
$\lambda=\lambda_2=1,~\lambda_3=8$ is presented in Figure 3. We have found that $|\varphi|$
possesses always two zeros at $\pm d/2$ on the $z-$symmetry axis,
the positions of the nodes depending on the value of the coupling constants $\lambda_i$.
The total action of these solutions, as given by the integral of
$E(R,\theta)$ increases with increasing $\lambda_i$.
%

%%%%%%%%%%%%%%%%%%%%%%%%
\begin{figure}[h!]
\parbox{\textwidth}
{\centerline{
\mbox{
\epsfysize=10.0cm
\includegraphics[width=87mm,angle=0,keepaspectratio]{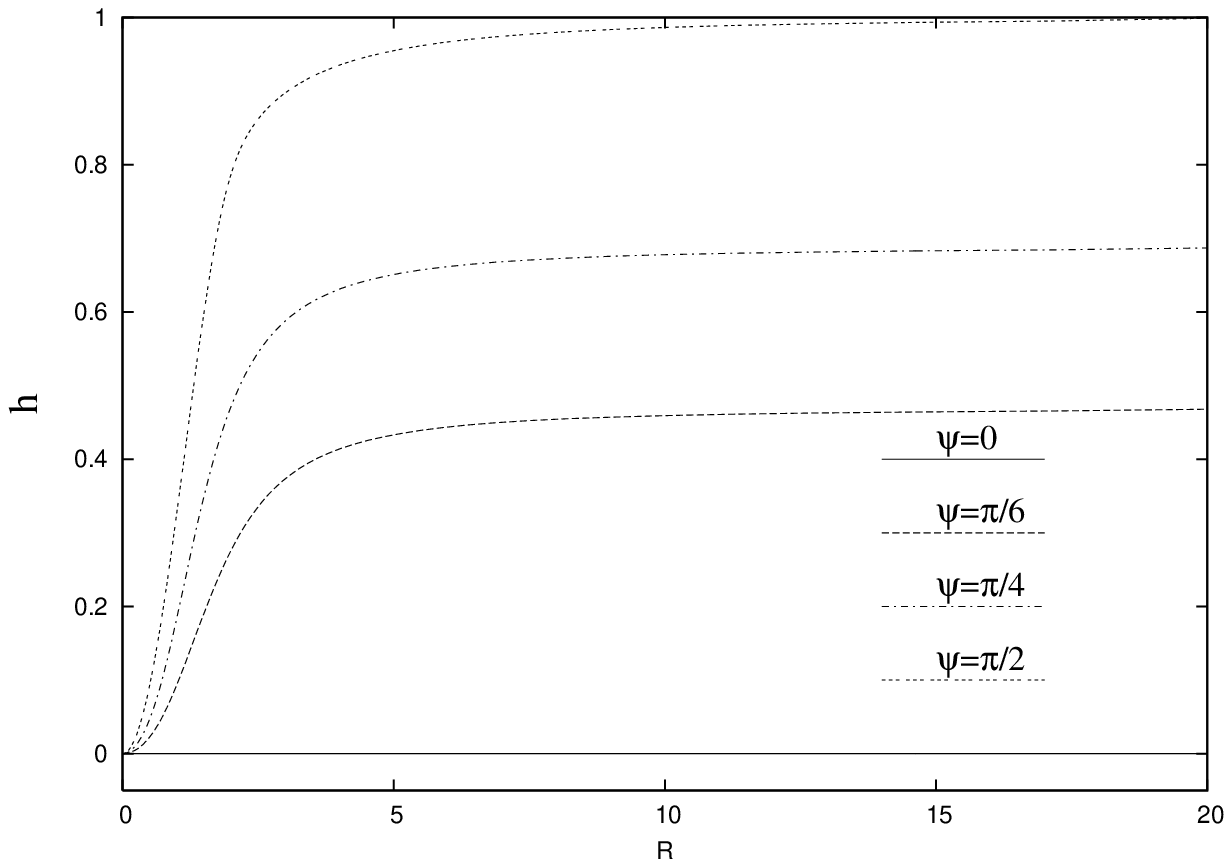}
\includegraphics[width=87mm,angle=0,keepaspectratio]{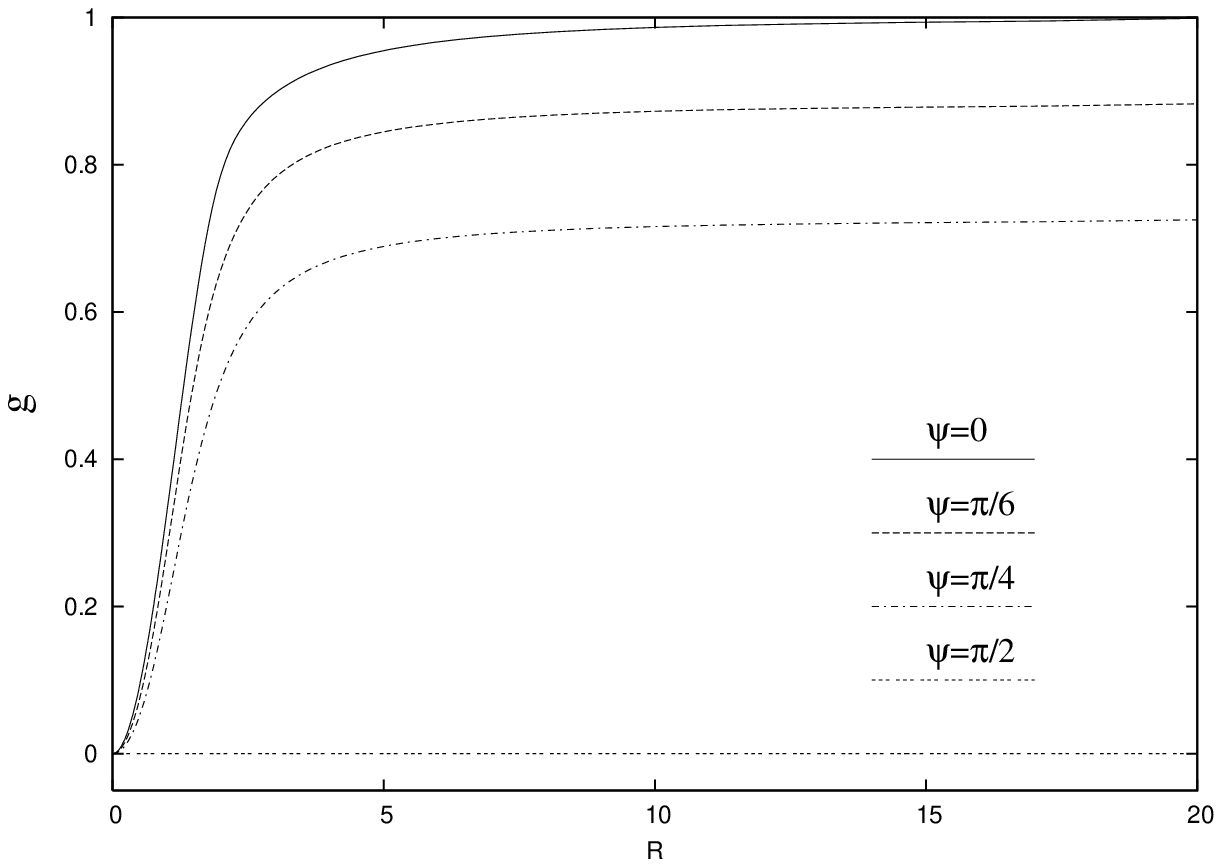}
}}}
\caption{{\small The profiles of the scalar functions 
$h$ and $g$ are shown for
a typical $D=4$ by-azimuthally symmetric  solution
with $n_1=n_2=2$, $\lambda_1=\lambda_2=\lambda_3=1$.}}
\end{figure}
%
%%%%%%%%%%%%%%%%%%%%%%%%%%

Interestingly enough, the weighted energy density  $E(R,\theta)$ possesses a saddle
point at the origin, the maxima being localized at $z=\pm d/2$, at a nonzero value of $r$, $r=r_0$.
This feature, already present in the $D=3$ case (see Figure 4 in Ref. \cite{Paturyan:2005ik}) 
is enhanced for the higher dimensional configurations, in contrast with the $D=3$
Yang-Mills--Higgs~\cite{KK,KKS} where the concentartions of energy are located exactly on
the symmetry axis.

Although the profiles of the axially symmetric solutions look qualitatively the same for
$D=3,~4,~5$, their physical significance is very different.
For $D=4$ they describe two distinct solitons sitting  
at ($z=\pm d/2$, $r=r_0)$, while in three and five dimensions  the solutions 
represent a pair of soliton-antisoliton with zero topological charge.

It would be interesting to construct higher $m$ solutions,
describing for an odd dimension soliton-anti-soliton chains,
in analogy with the situation in YMH theory \cite{KKS}.

%%%%%%%%%%%%%%%%%%%%%%%%%%%%%%%%%%%%%%%%%%%%%%%%%%%%%%%%%%%%%%%%%%%%%%%%%
\subsection{Solutions with bi-azimuthal symmetry}
%%%%%%%%%%%%%%%%%%%%%%%%%%%%%%%%%%%%%%%%%%%%%%%%%%%%%%%%%%%%%%%%%%%%%%%%%
To obtain $D=4$, $m=1$ configurations with bi-azimuthal symmetry, 
we employ the $n=1$ spherically symmetric solutions discussed in {\bf 5.1}
above for starting profiles and increase the values of $n_1,~n_2$ slowly.
The iterations converge, and repeating the procedure one obtains in this way solutions for arbitrary
$n$. The physical values of $n_1,~n_2$ are integers. We have studied solutions with
$1\leq n_1,n_2\leq 9$. The weighted energy density  $E(R,\psi)$ can be written 
in terms of the reduced building blocks (\ref{2b}).
The two scalar functions $h(R,\psi)$ and $g(R,\psi)$ satisfy the boundary conditions
\begin{eqnarray}
\label{BA-b1}
h|_{R=0}=g|_{R=0}=0
\end{eqnarray}
at the origin, (\ref{mh}) at infinity, and
\begin{eqnarray}
\label{sw2}
h|_{\psi=0}=\partial_{\psi}g|_{\psi=0}=0,
~~\partial_{\psi}h|_{\psi=\pi/2}=g|_{\psi=\pi/2}=0.
\end{eqnarray}
on the $\rho$ and $\sigma$ axes. The field equations have been solved by using the same methods employed
in the axially symmetric case but now with much better accuracy, the typical numerical error
being of the order of $10^{-4}$ or smaller.

%%%%%%%%%%%%%%%%%%%%%%%%
\begin{figure}[h!]
\parbox{\textwidth}
{\centerline{
\mbox{
\epsfysize=10.0cm
\includegraphics[width=102mm,angle=0,keepaspectratio]{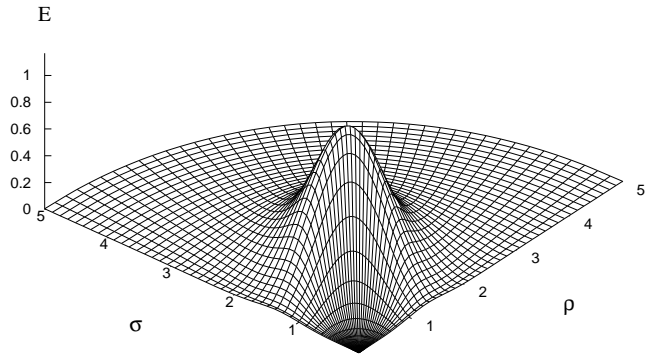} 
}}}
\caption{{\small 
A three-dimensional plot of the weighted energy density  
$E(\rho,\sigma)$ of the $D=4$ bi-azimuthally symmetric  solution
presented in Figure 4.}}
\end{figure}
%%%%%%%%%%%%%%%%%%%%%%%%%% 

As expected, the bi-azimuthally symmetric solutions exhibit a very different picture.
A general feature of all $m=1$ solutions with $n_1=n_2$ is that the weighted energy 
 density $E(R,\psi)$ possesses one maximum on the $\psi=\pi/4$ axis (corresponding to the 
$\rho=\sigma$ surface), it being possible to distinguish only one individual concentration of 
the action. In this respect the bi-azimuthally symmetric multisolitons of this model are 
qualitatively similar to those
of the models featuring gauge fields, as discussed recently in 
\cite{Radu:2006gg,Radu:2006qf}. In
contrast to the latter however, where only solutions with $n_1=n_2$ 
were found, here  
we noticed the existence
finite mass solutions with $n_1\neq n_2$.
The maximum of the weighted energy  density moves inward with increasing 
$n_1,n_2$.

In Figure 4 the profiles scalar functions $h$ and $g$ of the a 
typical bi-azimuthally symmetric
solution are shown for several angles as a function of the radial 
coordinate 
$R$ (with $\lambda_1=\lambda_2=\lambda_3=1$ in this case).
A three dimensional plot of the weighted energy density of a typical 
$m=1,~n_1=n_2=2$ configuration 
is presented in Figure 5. 
We have also studied the  mass dependence  of the bi-azimuthally 
symmetric solutions on the coupling constants $\lambda_i$. From the numerical results, we
observed some features of the solutions, without attempting to give an analytic explanation.
The most peculiar of these is the fact that finite mass solutions persist for a small
but finite range of negative $\la_2\ \le 0$.
No such solutions can be justified by the topological lower bounds. 
%
%
 
%
%%%%%%%%%%%%%%%%%%%%%%%%
\begin{figure}[h!]
\parbox{\textwidth}
{\centerline{
\mbox{
\epsfysize=10.0cm
\includegraphics[width=102mm,angle=0,keepaspectratio]{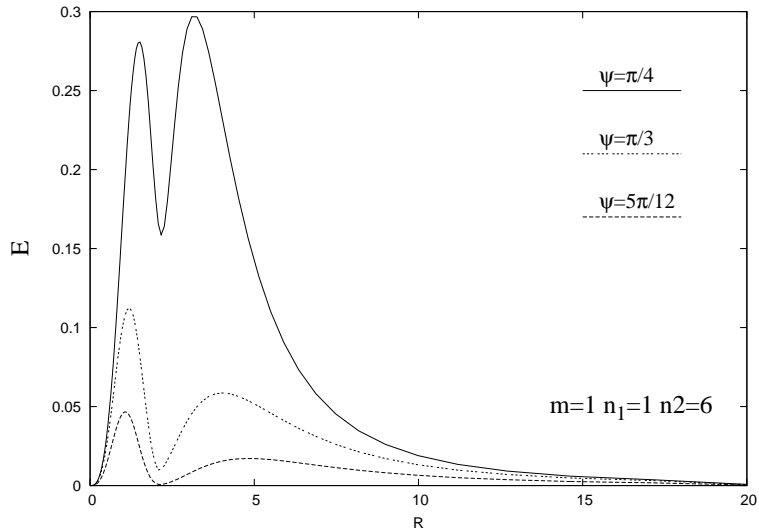} 
}}}
\caption{{\small 
The weighted energy density $E$ is shown for a $D=4$ bi-azimuthally symmetric  solution presenting
two localized  elementary constituents.
%is plotted as a function of the radial coordinate $R$ for several angles $\Psi$.
}
}
\end{figure}
%%%%%%%%%%%%%%%%%%%%%%%%%%
%
When $\lambda_1$ or $\lambda_2$ are varied, the maximum of the 
weighted energy  density moves inwards
with increasing value of the respective coupling constant, while the 
opposite behaviour to this
is found when $\lambda_3$ is varied. Yet another property of multi-solitons is observed. It turns out 
that the mass of a solution $M(n_1,n_2)$, whose topological lower bound given by \re{i4bi} to be
$4\pi^2\,n_1n_2$, is quantitatively quite close to $n_1n_2\,M(1,1)$. 
This means that the deviation of the value of $M(n_1,n_2)$ from its lower bound value is 
proportionate to the deviation of the value of $M(1,1)$ from its respective lower bound value, 
implying that composite solitons have rather low binding energies. On the other hand, in all cases 
studied, it turns out that $M(n_1,n_2)>n_1n_2\,M(1,1)$, albeit by a small amount quantitatively. 
This in turn suggests that the composite states are unstable against decay into the lowest mass 
constituents.

Yet another surprising but not counterintuitive property observed is, that while multi-solitons of
masses $M(n_1,n_2)$ exhibit one single peak of the weighted energy  density for {\it small} values of
$|n_1-n_2|$, when $|n_1-n_2|$ becomes {\it large} the weighted energy 
 density develops two separated peaks (see Figure 6).

More complicated bi-azimuthal solutions with $m\ge 2$ are likely to exist.
These configurations would describe composite bound states, rather analogous
to the monopole-antimonopole chains of the Yang-Mills-Higgs model \cite{KKS}.
Our preliminary numerical results already indicate the existence of zero topological charge
$m=2$ configurations with bi-azimuthal symmetry in $D=4$, with $n_1=n_2=2$. We found that there were
no $m=2$ solutions with $n_1=n_2=1$, just as for the pure YM model 
whose instanton-antiinstanton solutions constructed in \cite{Radu:2006gg}. Again as in 
\cite{Radu:2006gg}, the weighted energy density exhibits two distinct maxima on the $\psi=\pi/4$ axis.
In the absence of suitable starting profiles however, the numerical accuracy of these solutions turned 
out to be much lower in this case\footnote{We started the
numerical process by employing a guess solution  constructed
with a suitable trial functions that contains
several free parameters whose values are then tuned.
However, improving the accuracy of these profiles has proven  
more difficult than in the case of axially symmetric configurations, where
a similar approach has been employed.}.

We hope to return to a systematic discussion of these solutions, together with 
a generalisation to higher
dimensions.

%%%%%%%%%%%%%%%%%%%%%%%%%%%%%%%%%%%%%%%%%%%%%%%%%%%%%%%%%%%%%%%%%%%%%%%%%
\section{Summary and Conclusions}
%%%%%%%%%%%%%%%%%%%%%%%%%%%%%%%%%%%%%%%%%%%%%%%%%%%%%%%%%%%%%%%%%%%%%%%%%
The overriding aim of this work is to examine the conditions that enable the construction of
zero topological charge solutions in classical field theories, which otherwise support
topologically stable (multi-)soliton solutions with nonvanishing topological charge. We have shown
that the conditions in question are those of ($a$) subjecting the system to the requisite symmetry,
which in practice is what is done anyway when constructing (multi-)solitons,
and ($b$) by requiring special types of boundary conditions that differ essentially from those 
employed for (multi-)solitons.

Our symmetry analysis covers dimensions $D=3,4,5,6$, 
namely both odd and even examples in $6\ge D\ge 3$,
while the solutions constructed numerically to underpin 
our findings are limited to $D\le 5$. The reason
for this restriction is that boundary value 
problems in more than two dimensions is beyond the scope
of the present work.
The case $D=2$ is irrelevant, being too small to accommodate requirements ($a$) and ($b$).
The case $D=3$, while it is the first nontrivial example, is rather special since in that case
{\it axial} and {\it azimuthal} symmetries coincide. We have chosen to carry out this investigation
in the framework of the simplest possible field theoretic model, irrespective of its applicability
to physical problems. This is a symmetry breaking Goldstone type model 
in $D$ Euclidean dimensions,
whose energy density functional depends on a $D$ component scalar field $\f^a$,
$a=1,2,\dots,D$. This choice is motivated by the fact that the topological charges in such models are
the simplest available examples, being up to constant multiples the {\it winding numbers} of
$\f^a$ in $\R^D$. It should be emphasized however that our conclusions hold in the other classical
field theories that support solitons in these dimensions, 
namely the sigma models and the non Abelian
gauge field systems (including Higgs fields).

Our conclusions can be summarised as follows:
\begin{itemize}
\item
The field $\f^a$ asymptotically tends to a unit vector 
$\hat\f^a$, which depends exclusively
on angular variables, the radial variable being infinite.
 Precisely what these angular variables are
depends on the symmetry imposed. To analyse  qualitatively distinct possibilities, we have found it
sufficient to impose symmetries that result in residual subsystems of no more than {\bf three}
dimensions. It is superfluous to consider weaker symmetries resulting in four or higher number of
effective degrees of freedom, since these do not result in qualitatively new features as far as the
existence of multisolitons, and soliton--antisoliton chains is concerned.

\item
There are two types of symmetries employed. First, spherical (rotational) symmetry in a $N$ dimensional subspace
of $\R^D$. For $N=D-1$, this is {\it axial} symmetry resulting in {\it two} effective degrees of freedom.
At the other extreme, $N=2$, this is {\it azimuthal} symmetry resulting in an effective $D-2$ dimensional
subsystem. Accordingly, we have restricted to $D=3,~4$ in the case 
of {\it azimuthal} symmetry.  Intermediate
values of $N$ subject to this restriction are, $N=3$ in $D=5$ and $N=4$ in $D=6$. We have described these
as {\it intermediate} symmetries. Second, we impose {\it multi-azimuthal} symmetries, composed of
azimuthal symmetries in pairs of coordinates. Again, subject to 
limiting our considerations to three effective
degrees of freedom, these are {\it bi-azimuthal} symmetry in $D=4,5$, and {\it tri-azimuthal} symmetry in $D=6$.
The residual system after symmetry imposition depends on the radial coordinate and the remaining angular
coordinates, and the number of unknown functions is the same as the dimensionality of the residual subsystem.
In all cases, the azimuthal angles are all integrated out and the remaining angular dependence is on
{\it polar} angles $\{\ta_i\}$ ($0\le\ta\le\pi$) and {\it semi-polar} angles $\{\psi_I\}$
($0\le\psi_I\le\frac{\pi}{2}$), $i$ and $I$ labeling the residual polar and semi-polar angles, respectively.
\item
The asymptotic field $\hat\f^a$ is parametrised by the residual angular variables $\{\ta_i\}$ and $\{\psi_I\}$
only. The other angular variables, that include {\bf all} {\it azimuthal} angles, are integrated out. Consistently
with the requirements of finite energy and analyticity, the most general as $\hat\f^a$ is encoded by $\{\ta_i\}$
and $\{\psi_I\}$ is via
\be
\label{m}
\{m_i\,\ta_i\}\quad,\quad\{m_I\,\psi_I\}\quad,\quad m_i\,,m_I\quad{\rm integers}\,.
\ee
It is important to stress that the integers $(m_i\,,m_I)$ in \re{m} appear {\bf only} 
the in the asymptotic field
$\hat\f^a$, and that they {\bf do not} parametrise the field $\f^a$ everywhere. 
Throughout this text, we have
reserved the letter $m$ to these $m$--numbers. In contrast we label the {\it vorticity} 
associated with
each azimuthal symmetry, on which the field $\f^a$ everywhere depends, by the letters $\{n\}$.
Thus the $n$--numbers which count the winding in each azimuthal plane are on a completely distinct footing as 
opposed to the $m$--numbers which serve only to select the boundary values imposed. All solutions with
$m=1$ describe topologically stable~\footnote{These solutions are not {\it absolutely} stable since no first
order Bogomol'nyi type are saturated in these models. Rather, the stability in question is a consequence of
the energy respecting topological lower bound, like in the case of Skyrmions~\cite{Manton:2004tk}}
multisolitons whose topological charges are encoded with the 
$n$--numbers, $\{n\}$. The topological charges of soliton--antisoliton chains with even $m$ are {\it zero}, 
while those of odd $m$ are {\it nonzero}, and depend on $\{n\}$. In the special case of axial symmetry 
in $D\ge 4$, when no $n$--number occurs, the topological charges in even $D$ are labeled by a $m$-number.
\end{itemize}
The numerical constructions in Section {\bf 5} underpin the above conclusions.
Both (multi)solitons and soliton--antisoliton solutions have been constructed like for the $D=3$ case in
\cite{Paturyan:2005ik}, whose results are extended to higher dimensions the present work.
Various features found there are shared by higher dimensional axially symmetric
solutions. In particular the profiles of the scalar functions have rather similar shapes.
The numerical constructions in the present work is limited to $2$ dimensional boundary value problem
involving two functions, so that we present only axially symmetric solutions in $D=4,5$ (and $D=3$ in
\cite{Paturyan:2005ik}), and bi-azimuthally symmetric solutions in $D=4$.

The axially symmetric solutions constructed in both \cite{Paturyan:2005ik} and in section {\bf 5.2} here are
limited in their scope to solutions with asymptotic behaviour characterised by $m-$number equal to
$1$ and to $2$. In the $D=3$ case~\cite{Paturyan:2005ik} these are multisolitons with $m=1$ and higher $n-$numbers,
and to soliton--antisoliton pairs with $m=2$ and $n-$number equal to $1$. 
In $D=3$, like for the YMH monopoles~(see e.g. \cite{Manton:2004tk} and \cite{Shnir:2005xx})
the energy density of the multisolitons is concentrated at the origin, and when $m=2$ 
there occur two concentrations
distributed symmetrically on the $3-$axis. While we expect that solitons with $m\ge 2$ 
and with $n=1$ would
describe chains of solitons and antisolitons on the $3-$axis, and when $n\ge 3$ rings would form,
like for the monopoles in the YMH model observed in  \cite{KKS}, this has not 
been carried out in \cite{Paturyan:2005ik}.
Here, axially symmetric solutions in $D=4,5$
are constructed in section {\bf 5.2}. The $m=2$ solutions in $D=5$ are similar to the $m=2\ ,\ n=1$
soliton--antisoliton solutions in $D=3$, {\it i.e.} they describe two distinct 
peaks of the weighted energy density on the $5-$axis
of equal and opposite charges. But there is no $n-$number in $D=5$ so here the 
analogy with $D=3$ stops.
The $m=2$ axially symmetric solutions in $D=4$ also describe two distinct particles
located on the $4-$axis, 
but unlike those of $D=3,5$
they both peaks of the energy density 
have the same topological charge. These are multisolitons, qualitatively 
similar to the axially symmetric
Witten multiinstantons~\cite{Witten:1976ck}. This illustrates that for axially 
symmetric fields with $m-$number higher than $1$,
the peaks of the weighted energy density 
are situated on the (symmetry) $D-$axis, such that in odd dimensions 
their charges have alternating signs,
while in even dimensions all the charges have the same sign.

There is one final property of axially symmetric solutions worth remarking on.
In gauge field systems, namely the $D=3$ YMH configurations~
as our only example, the zero charge $m=2$
solutions have a positive binding energy with respect to decay into two charge$-1$ monopoles
 \cite{KK}. 
By contrast,
the multisolitons and the soliton--antisoliton solutions of the Goldstone models have negative 
binding energies.
Because the data available to us is not sharp enough,
we have not displayed this quantitatively either with a plot or a table in section {\bf 5.2}.
Nevertheless the observed qualitative trend is unmistakable.

The bi-azimuthally symmetric solutions in $D=4$ constructed in section {\bf 5.3} have their 
analogue in
the bi-azimuthal YM instantons given in \cite{Radu:2006gg}. Like in that case there is 
a $n-$number associated
with each (of the two) azimuthal symmetries, $n_1$ and $n_2$, and the topological charge 
is proportional to
$n_1\,n_2$ (see \re{i5bi}). Unlike in \cite{Radu:2006gg} however, where finite action
solutions occur only for one interger $n_1=n_2=n$, here there are solutions for 
distinct $n_1\neq n_2$.
For $m-$number equal to $1$ with $n_1=n_2=n$ the action density has only one peak 
which like in the YM
example~\cite{Radu:2006gg} is not situated at the origin. Rather, it peaks at a 
numerically determined
distance from the origin on the $\psi=\frac{\pi}{4}$ axis, $\psi$ being the unique 
semi--polar angle. The
situation is different in the ($m=1$) $n_1\neq n_2$ case. There the action density 
breaks up into two distinct
peaks on the $\psi=\frac{\pi}{4}$ axis, and the centres of these peaks move away from each other as
$|n_1-n_2|$ increases. Another point of contrast with the YM case, where the multiinstantons 
do form bound
states, the corresponding multisolitons here do not form bound states. 
For all configurations we have studied,
the energy of a  the $n_1,~n_2$ multi--solitons of our model are greater than that of $n_1n_2$ $1$-solitons. 
Moreover it turns out that this deficit of binding energy increases with increasing $n_1,~n_2$.
We have also verified that $m=2$ solutions carrying zero topological charge
(see \re{i5bi}) exist, in the $n_1=n_2=2$ case, but have not supplied quantitative data here. 
These present two
distinct peaks of the weighted energy density
like in the YM case~\cite{Radu:2006gg}. Our numerical results here were not sufficiently
accurate to enable us to estimate wheter the binding energy preventing the decay of this solution into
two charge$-2$ ($m=1$) multisolitons is positive or negative. Likewise for the same reason, we did not
increase $n_1$ and $n_2$ to values higher than $3$, to see what the analogues of the rings forming in
the $D=3$ YM example~\cite{KK} are. (Such ring like configurations were discovered recently in the
bi-azimuthal gauge field configurations with $n=3$ in \cite{Radu:2006qf}, implying their ocurrence here.)

%Finally, we mention that, as a new type of configurations, we have also presented numerical evidence for the existence of $D=4$ multi--soliton solutions with bi-azimuthal symmetry. An interesting feature of these solutions is the absence of bound states.

This completes the summary of our results. We now make some final, general comments. We have seen that
most of the geometrical and topological properties of the multisoliton and soliton--antisoliton solutions
in the Goldstone models studied here, are broadly similar both to the YMH example~\cite{KK,KKS} in $D=3$ and to
the YM example~\cite{Radu:2006gg} in $D=4$. There are however some notable differences, firstly that the binding
energies of our multisolitons are negative as opposed to those of their gauge field
counterparts~\cite{KK,KKS,Radu:2006gg}, which are positive. 
Then there is the difference between the $D=4$
bi-azimuthal Goldstone solitons, where the two vorticities 
($n-$numbres) can be different, and the $D=4$
YM instantons for which the two vorticities must be equal. 
More recently the $SU(2)$ YM-dilaton system in $4+1$
dimensions was analysed and the static bi-azimuthally symmetric 
solutions were studied in \cite{Radu:2006qf}.
There too, the numerical results indicated that the two $n-$numbers 
had to be equal $n_1=n_2$. It appears
therefore that this restriction ($n_1=n_2$) applies to 
bi-azimuthally symmetric gauge fields, but not
to Goldstone fields. It is likely this feature may persist 
in multi-azimuthal systems too, but since this conclusion
is reached only on the basis of numerics, it is beyond the scope of the present work.

Based on what we have learnt about the general similarities
in the different models supporting topologically nontrivial lumps studied 
here and in \cite{KK,KKS,Radu:2006gg,Radu:2006qf},
we would speculate that similar analogous properties can be expected for 
the lump solutions in various sigma models,
{\it e,g.} $O(D+1)$ models on $\R^D$, or the corresponding Grassmannian sigma 
models on $\R^{2N}$, or indeed their
gauged counterparts. One respect in which it would have been more appropriate 
to use $O(D+1)$ models on $\R^D$ instead
of Goldstone models, featuring negative binding energies, is that the $O(D+1)$ 
models would be expected to feature positive
binding energies, based on our knowledge of the $O(3+1)$ model 
on $\R^3$, namely the celebrated Skyrme model.
Certainly the simple analysis of the topological charges and boundary conditions 
given in section {\bf 4} can
be extended systematically and without obstacles to the sigma 
model counterparts of the scalar Goldstone fields.
This was eschewed because the numerical constructions for the 
sigma models, in particular the practical task of
imposing the boundary conditions, are very much harder.

One last comment concerns a common feature of zero topological 
charge bi-azimuthal solutions to both gauge field systems,
namely those studied in \cite{Radu:2006gg,Radu:2006qf}, and to the correspoding 
Goldstone model studied here. These
are both solutions with $m-$number equal to $2$. In the former case, 
the numerical results indicated that the simplest
such solution was that with  $n-$number equal to $2$, and not $n=1$. 
Likewise in the case at hand, it turned out that
there existed no solution for $n_1=n_2=1$, the simplest solution being 
characterised by $n_1=1\ ,\ n_2=2$. It is
interesting that this observation is consistent with the results of the numerical 
analysis of Krusch and
Sutcliffe~\cite{Krusch:2004uf} in the context of the zero baryon charge solutions 
of the Skyrme model. It is very
interesting also that the analytic analysis of Sadun and Segert~\cite{SS2}, 
which proves the existence of non--selfdual instantons
({\it e.g.} the $m=3$ instantons in~\cite{Radu:2006gg}), their proof excludes 
topological charge $1$ instantons
({\it e.g.} our $m=3$ instantons with $n=1$ whose topological charge is
 equal to $n^2=1$). This is a rather subtle
but pervasive feature, which we cannot analyse further here.

\medskip

%\newpage
\noindent
{\bf Acknowledgement}
\\
This work was carried out in the framework of Science Foundation Ireland
(SFI) Research Frontiers Programme (RFP) project RFP07/FPHY330.

\end{document}